%% file: main.tex
\documentclass{sig-alternate-10pt}
\usepackage{graphics}

\usepackage{booktabs,tabularx}

\usepackage{graphicx}
\usepackage{comment}
\usepackage{xspace}
\usepackage{balance}
\usepackage{subfigure}
\usepackage{url}
\usepackage{color}
\usepackage{multirow}

\usepackage[binary-units = true]{siunitx}
\usepackage{mdwlist}

\usepackage{amsmath}
\usepackage{amssymb}
\usepackage[latin1]{inputenc}
\usepackage[T1]{fontenc}

\usepackage{mwe}

\usepackage[linesnumbered,ruled,vlined,onelanguage]{algorithm2e}
\usepackage{flexisym}

\newcommand{\fakeparagraph}[1]{\vspace{.5mm}\noindent\textbf{#1.}}
\newcommand{\fakepar}[1]{\fakeparagraph{#1}}
\newcommand{\capt}[1]{\mdseries{\emph{#1}}}

\newcommand{\highspeed}{\textsc{High-speed}\xspace}
\newcommand{\hgain}{\textsc{High-gain}\xspace}
\newcommand{\lgain}{\textsc{Low-gain}\xspace}
\newcommand{\ulp}{\textsc{Ultra-low-po\-wer}\xspace}
\newcommand{\platform}{\textsc{Integration}\xspace}

\newfont{\mycrnotice}{ptmr8t at 7pt}
\newfont{\myconfname}{ptmri8t at 7pt}
\let\crnotice\mycrnotice%
\let\confname\myconfname%

\begin{document}

\title{Visible Light Communication for Wearable Computing}

\author{
%
\numberofauthors{3}
\alignauthor
Ambuj Varshney\\
       \affaddr{Uppsala University, Sweden}\\
       \email{ambuj.varshney@it.uu.se}
\alignauthor
       Luca Mottola \\
       \affaddr{Politecnico di Milano, Italy and RISE SICS AB, Sweden}\\
       \email{luca@sics.se}
\and  
\alignauthor Thiemo Voigt\\
       \affaddr{Uppsala University, Sweden  RISE SICS AB, Sweden}\\
       \email{thiemo@sics.se}
}

\maketitle

\input{abstract}

\input{intro}

\input{background}

\input{design}

\input{platform}
\input{eval}

\input{usecase}
\input{end2.tex}

\bibliographystyle{abbrv}
\bibliography{ref}
 
\end{document}

%% file: abstract.tex
\begin{abstract}

Visible Light Communication (VLC) is emerging as a means to network computing devices that ameliorates many hurdles of radio-frequency (RF) communications, for example, the limited available spectrum. 
Enabling VLC in wearable computing, however, is challenging because mobility induces \emph{unpredictable drastic changes} in light conditions, for example, due to reflective surfaces and obstacles casting shadows.
We experimentally demonstrate that such changes are so extreme that no single design of a VLC receiver can provide efficient performance across the board. 
The diversity found in current werable devices complicates matters. 
Based on these observations, we present three different designs of VLC receivers that \emph{i)} are individually orders of magnitude more efficient than the state-of-the-art in a subset of the possible conditions, and \emph{ii)} can be be combined in a single unit that dynamically switches to the best performing receiver based on the light conditions.
Our evaluation indicates that dynamic switching incurs minimal overhead, that we can obtain throughput in the order of MBit/s, and at energy costs lower than many RF devices.


\end{abstract}


%% file: intro.tex
\section{Introduction}

Visual Light Communication (VLC)~\cite{1277847,6163585} is rapidly emerging as a complement or alternative to ra\-dio-fre\-quency (RF) communications.
VLC, in fact, plays host to several unique features.
For example, it provides a wide spectrum available for communication orthogonal to that of RF communications, thus reducing cross-technology interference, and does not penetrate walls, which facilities spatial re-use and offers a means to physically secure data transfers.
Novel networking abstractions, such as room-area networks~\cite{iannucci2015room}, are implemented straightforwardly with VLC. 

Wearable computing is similarly gaining momentum.
Wearable devices ranging from smart-watches~\cite{applewatch} to embedded sensors harvesting energy from the human bo\-dy~\cite{Bhatti:2016:EHW:2976745.2915918,Gorlatova:2014:MSK:2591971.2591986} enable a range of novel applications~\cite{billinghurst1999wearable}.
It may appear natural that the two technologies are destined for a happy marriage.
For example, the omnipresent Light Emitting Diodes (LEDs) installed in modern buildings may use VLC to provide context and configuration information to wearable devices without impacting other co-located wireless transmissions.
Provided proper permissions are granted by the owner, LEDs in a building may even be used to deploy dedicated apps on a wearable device as soon as one enters the building; for example, to control HVAC systems therein.  

\begin{figure}[!tb]
\centering
\includegraphics[width=0.8\linewidth]{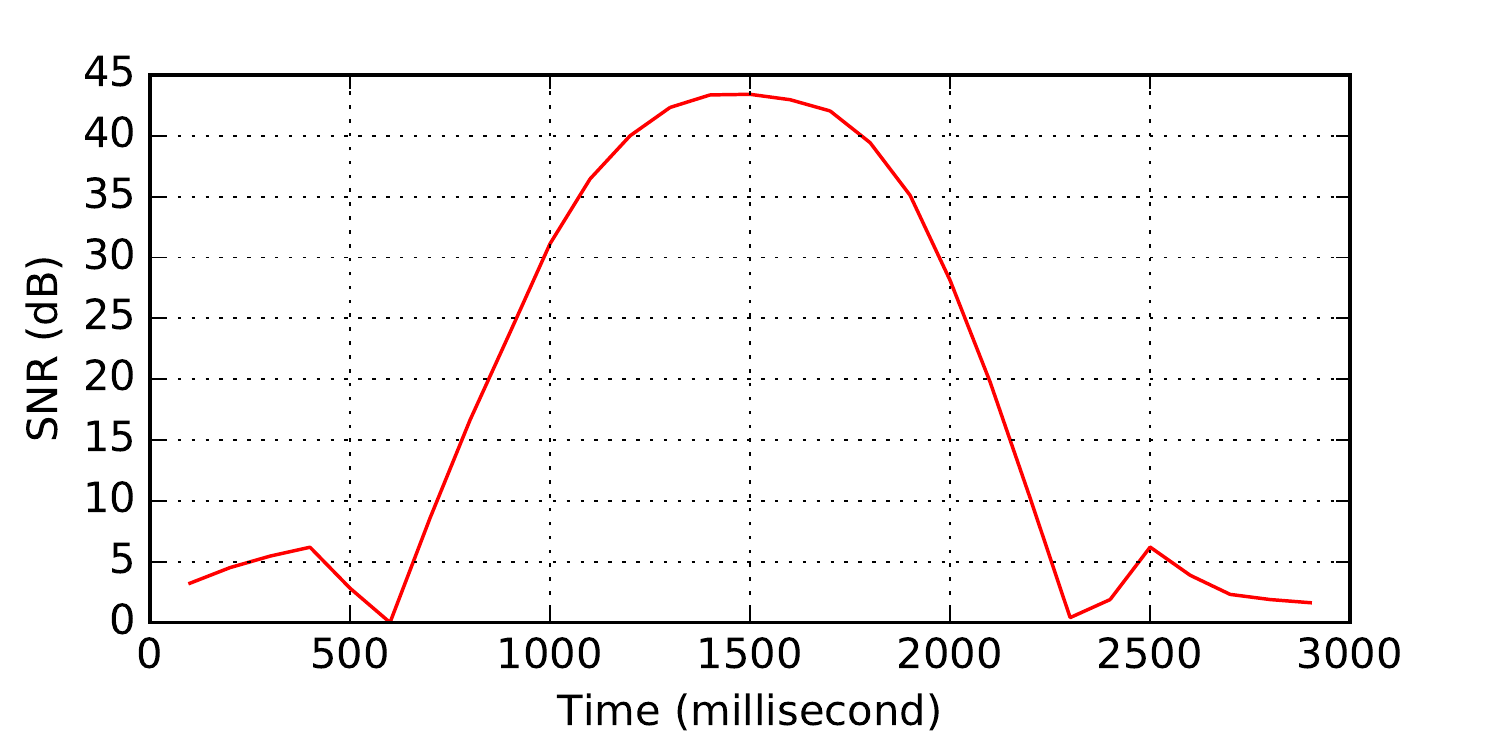}
\caption{Change in signal-to-noise (SNR) ratio under mobility. \capt{Mobility induces rapid changes in the SNR in a span of hundreds of milliseconds. No single VLC receiver design can provide efficient performance across the whole range.}}
\label{fig:snr_motivation}
\end{figure}

\fakepar{Challenge} Such a happy marriage, however, is fundamentally hampered by two key features of wearable computing, that are, \emph{mobility} and \emph{diversity}.


The efficiency of VLC receivers is a function of the signal-to-noise ratio (SNR) between the modulated light carrying the data and the surrounding light conditions.
Factors such as reflecting surfaces and obstacles that cast shadows, combined with a person's mobility, expose different parts of the body to drastically different light conditions.
The resulting changes in SNR are unpredictable and drastic~\cite{zhang2015dancing}.
Figure~\ref{fig:snr_motivation}, which we experimentally obtained by checking the light conditions on a person's chest when \emph{walking} in a university building, demonstrate that the SNR may change by tens of~dB in a few hundreds of milliseconds.

To complicate matters, devices for wearable computing exhibit an impressive degree of diversity in size, capabilities, and intended mode of use.
For example, today's smart-watches are powerful computing machines able to run multiple applications in parallel with rich graphical user interfaces.
Although battery-powered, users are supposed to re-charge the devices periodically. To complicate matters further, these devices also exhibit very different battery sizes and capacity~\cite{Braidio}. Sensors embedded in fabric, on the other hand, are extremely constrained computing machines that may be solely powered by ambient energy if and when available.
Energy preservation thus becomes paramount.

With current technology, we argue it is prohibitively difficult, if not impossible, to design a VLC receiver device able to: \emph{i)} seamlessly work in a range of extremely different light conditions, to the extent of causing the SNR changes in Figure~\ref{fig:snr_motivation}, and \emph{ii)} to simultaneously address the needs of powerful devices such as smart-watches as well as those of energy-constrained wearables such as embedded sensors.
The problem in \emph{i)} is particularly acute; the method of light sampling and gain configurations are necessarily optimized for a relatively narrow range of light conditions, which prevents them from working when such conditions vary wildly. 

\fakepar{Contribution} Rather than attacking the problem as a whole, we design \emph{three} different VLC receivers that individually address a subset of the issues at stake.
Next, we show that the receivers can be combined in a single unit that dynamically switches to the best performing receiver depending on the light conditions.

Following a survey of existing efforts in the field reported in Section~\ref{sec:background}, Section~\ref{design} describes the design of the individual VLC receivers.
Each within its own slice of the problem space, these designs offers performance orders of magnitude better than state-of-the-art VLC receivers.
The \ulp receiver departs from existing VLC designs by employing a solar cell as light sensor, achieving unprecedented energy efficiency whenever the light levels fall within appropriate operating range.
It also employs a thresholding circuit to digitize the signal that, unlike traditional designs, autonomously adapts to changes in the rapidly varying light conditions induced by mobility.
The \hgain and \lgain receivers---actually instantiated from the sa\-me fundamental design---also employ a thresholding circuit to gain the same benefits under mobility and, unlike existing VLC designs, can be configured to optimize throughput while avoiding saturation effects.

\begin{figure}[!tb]
\centering
\includegraphics[width=0.8\linewidth]{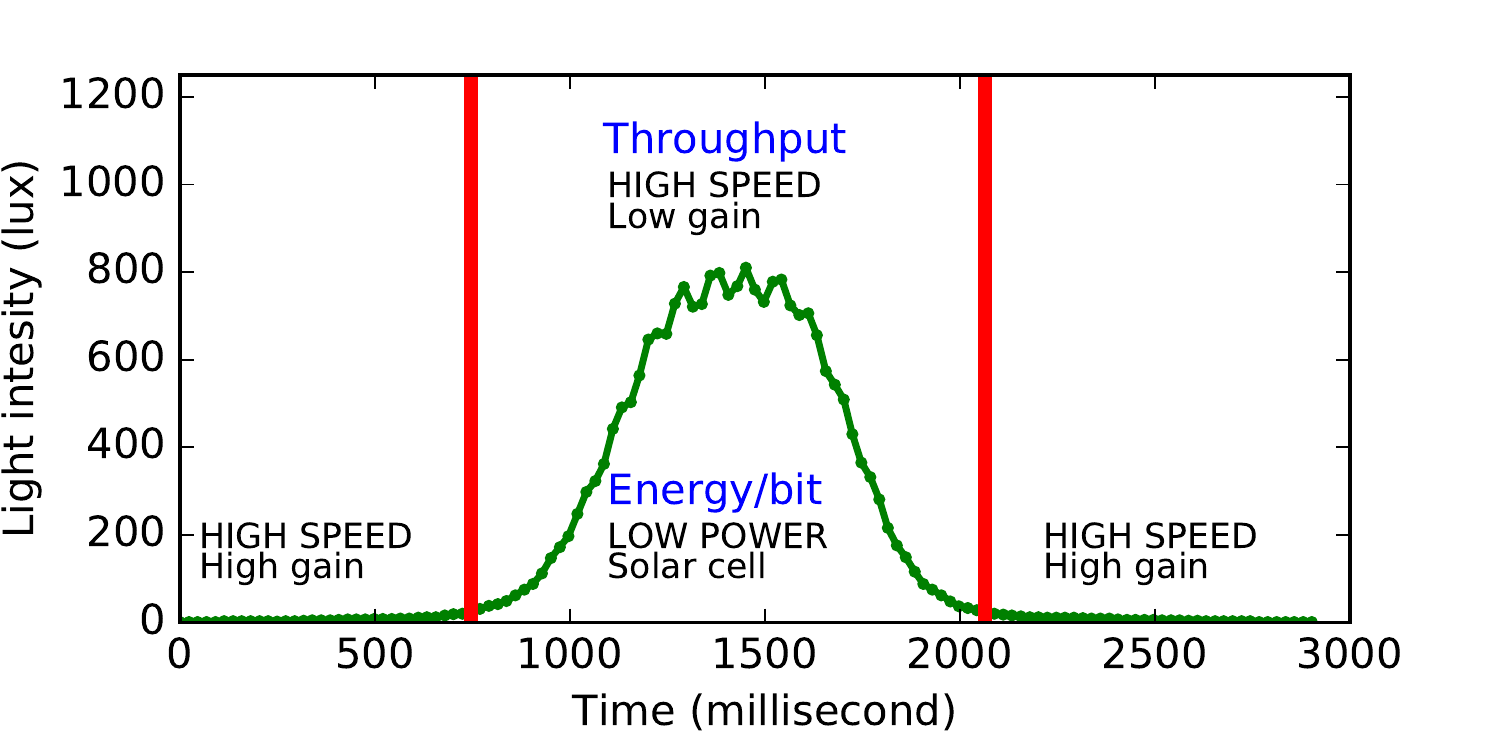}
\caption{An intuitive illustration of how to switch between the three VLC receivers depending on performance goals and light conditions.}
\label{fig:motivation}
\end{figure}

Section~\ref{sec:platform} describes the logic to dynamically switch among the three receivers, whose functioning is intuitively shown in Figure~\ref{fig:motivation}.
Different performance goals correspond to a different switching logic.
Figure~\ref{fig:motivation} shows that when optimizing energy per bit, in bright light conditions the \ulp receiver offers the best performance; otherwise, the \hgain receiver allows one to pick up even very small light emissions and still decode data at reasonable energy costs.
The latter applies also when optimizing for throughput, as shown in Figure~\ref{fig:motivation}; however, in bright light conditions the \hgain receiver saturates, which makes switching to the \lgain receiver the best choice, allowing one to transfer data up to MBit/s. 

Such a switching logic can be hosted at the wearable device, which attaches directly to the three receivers.
Resource-constrained devices may, however, not have sufficient resources.
Still in Section~\ref{sec:platform}, we describe the design of an energy-efficient integration unit dedicated to run the switching logic on behalf of the wearable devices, simplifying the integration and alleviating the processing requirements at the latter.

The experimental evaluation we carry out, reported in Section~\ref{sec:eval}, shows that our individual VLC receivers perform orders of magnitude better than state-of-the-art VLC receivers: the \ulp receiver's energy consumption is lower than even those of low-power RF chips, and in fact similar to the receivers employed on passive backscatter tags~\cite{ambientbackscatter}, while the \highspeed receiver's throughput (and energy consumption) is on par with that of WiFi and Bluetooth. 
Moreover, our switching logic that we implement on an integration unit together with the VLC receivers, is able to dynamically select the best performing receiver rapidly when the light conditions change which enables low bit errors rates even in mobile settings where the light conditions vary drastically.
We present concluding remarks in Section~\ref{sec:end}.

%% file: background.tex
\section{Related Work}
\label{sec:background}

VLC in wearable computing represents largely un-explored territory.
Significant work exists, however, in high-throughput VLC, embedded VLC, visible light-based sensing, and body area networks. 

\fakepar{VLC and wearable computing}  Existing attempts to leverage VLC to network wearable devices have been largely limited to using smartphones or other devices equipped with cameras. Yang et al. develop a system that by changing the polarisation of the light enables any light source to 
communicate to resource constrained mobile devices equipped with cameras~\cite{wearableVLC}. Lee et al. exploit the rolling shutter effect on cameras to enable communication between LEDs and smartphones, achieving a throughput of 11.32 $bps$~\cite{RollingLight}. Hu et al. improve upon that work using color shift keying~\cite{ColorShift}. They achieve a throughput of 5.2 $kbps$. 
	All of these systems require an energy-expensive camera,  extensive processing and achieve low throughput. This makes
	it difficult to network wearable devices, in particular, wearable devices that operate only on harvested energy.


\fakepar{High-throughput VLC} Whenever throughput is the only performance metric, VLC can deliver remarkable performance~\cite{1649137}.
For example,  Tsonev et al.\ demonstrate  data rates exceeding 3 $GBit/s$~\cite{6701327}. 
They use specialized  photodiodes coupled with high-speed ADCs or oscilloscopes to 
receive transmissions.

Zhang et al.\ point out that VLC channels are highly deterministic for given node locations and orientations~\cite{zhang2015dancing}.  They show, however, that small changes in movement or orientation can lead to SNR variations of tens of dB even in very short time spans.
We observe similar trends as shown in Figure~\ref{fig:snr_motivation}.
To cope with these issues, Zhang et al.\ present an in-frame rate adaption scheme.

VLC systems predominantly use photodiodes as light sensors~\cite{pathak2015visible}.
Wang et al.\ demonstrate the use of a solar panel as a light sensor, coupled to an energy-expensive voltage amplifier~\cite{wang2015design}.
They achieve 11.84 $MBit/s$ using OFDM as a modulation scheme.

The complexity and high energy costs make such solutions unsuitable for wearable computing. 
As a comparison, our \ulp receiver does not need voltage  amplifiers and ensures little processing overhead for constrained energy harvesting wearable devices.

\fakepar{Embedded VLC}
Recent efforts to connect embedded devices using VLC achieve much lower throughput.
Tian et al.\ present a VLC system working at light levels imperceptible to human eyes, achieving a throughput of 1.6 ~$kbps$~\cite{darkVLC}. 
Schmid et al.\ report a maximum throughput of 800 $bps$ while absorbing 288 $mW$~\cite{ledtoled}.
Our VLC receivers individually achieve both much higher throughput and lower energy consumption in a subset of the possible light conditions.
However, the ability to dynamically switch among them allows us to constantly reap the most benefits across the board.

Li et al.'s Retro-VLC enables full-duplex VLC~\cite{Li:2015:REB:2699343.2699354}.  
They achieve 10 $kbps$ of  throughput at an energy cost of \SI{125}{\micro\watt}.
To achieve low energy consumption, Li et al.\ use a thresholding circuit, like we do in the \ulp receiver.
However, while we use a solar cell as light sensor, Retro-VLC employs a photodiode together with an amplifier which  increases power consumption.
As a comparison, our \ulp receiver achieves 60 $kbps$ at a power consumption of \SI{0.5}{\micro\watt}, which represents a 250x improvement.

Our work is related to the design of a visible light receiver presented by Wang et al.~\cite{jsac} that adapts to the dynamics of visible light by switching between an LED and a photo-diode. Our design is similar, however, we switch between two complementary gain photodiodes to tackle the dynamics of visible light caused by mobility in wearable computing.  
We design an energy-efficient receiver that   significantly improves the throughput when compared to  their design.  Further, the thresholding circuit enables us to dynamically and in hardware calculate the threshold required to detect the high and the low symbols, which enables us to sustain a high throughput even under the dynamics of  visible light. As future work,  we will explore the use of an LED as receiver to support operation in very bright light conditions. 

Our work is also related to Purple VLC~\cite{purplevlc}, and OpenVLC 1.2~\cite{openvlc12} which use the Programmable Runtime Unit~(PRU) on the Beaglebone platform to support throughput as high as 100 KBit/s. We present a receiver design that achieves an order of magnitude higher maximum throughput, and our ULTRA-LOW-POWER receiver can achieve comparable throughput while consuming  three orders of magnitude lower power consumption. Further, PRUs are only available on certain platforms such as the BeagleBone. We demonstrate that our receiver can integrate with UARTs that are available on  the majority of MCU which brings high-speed VLC to a vast number of embedded devices.

Energy concerns are extreme for devices that harvest energy, for example, RF transmissions~\cite{ambientbackscatter}, like the Moo~\cite{zhang2011moo} and WISP~\cite{sample2008wisp} platforms.
The energy-hungry components traditionally used in VLC do not play along their characteristics.
This is precisely why very little work exists to network this kind of devices using VLC.
We demonstrate, however, that the performance of our \ulp receiver makes VLC applicable even for this class of wearable devices.

 \fakepar{Visible light-based sensing} VLC is also applied for sensing applications, such as indoor localization and gesture recognition.
Epsilon~\cite{epsilon} uses fixed LEDs to broadcast beacons to embedded device leading to submeter localization accuracy in office environments.
Rajagopal et al.\ present a VLC-based system that communicates to mobile receivers and embedded tags to enable room level localization~\cite{Rajagopal:2014:VLL:2602339.2602367}.
Li et al.\ describe a system that uses an array of photodiodes to  track user gestures by detecting the shadow they cast~\cite{Li:2015:HSU:2789168.2790110,practicalhumansensing}. Zhang et al.'s system uses light sensors to track the movement of fingers for accurate gesture detection~\cite{Okulli}.
Similar to us, these systems use off-the-shelf components together with embedded platforms as receivers.
Our work, however, targets mobile devices where links 
\emph{drastically} 
change even within short time spans.

Finally, we build on our prior work that introduces a solar cell coupled with a thresholding circuit to support ultra-low power visible light sensing. We extend the work to VLC, and demonstrate that this enables VLC on energy-constrained  and battery-free devices such as the WISP. Further, we also develop a high-speed receiver that achieves significantly higher throughput, and can operate under the dynamics of changing visible light.


\fakepar{Body area networks} 
Traditional body area networking employs RF transmissions to exchange data among on-body devices or between these and an external infrastructure~\cite{chen2011body}. 
Recent work in the area includes mechanisms that enable commodity input devices, such as those on wearable devices, to communicate securely using the body as a communication channel~\cite{hessar2016enabling}.
Moreover, Zhang et al.\ devise a low power mechanism to backscatter and shift ambient wireless signals to network sensors on wearable devices~\cite{zhang2016enabling}.

We see VLC as an addition to the toolbox of body area networks that provides key advantages over RF, such as reducing cross-technology interference,
when sending data from an infrastructure to on-body sensors. 
Our work provides the enabling technology to this end.


%% file: design.tex
\begin{figure}[!tb]
\centering
  \includegraphics[width=0.5\textwidth]{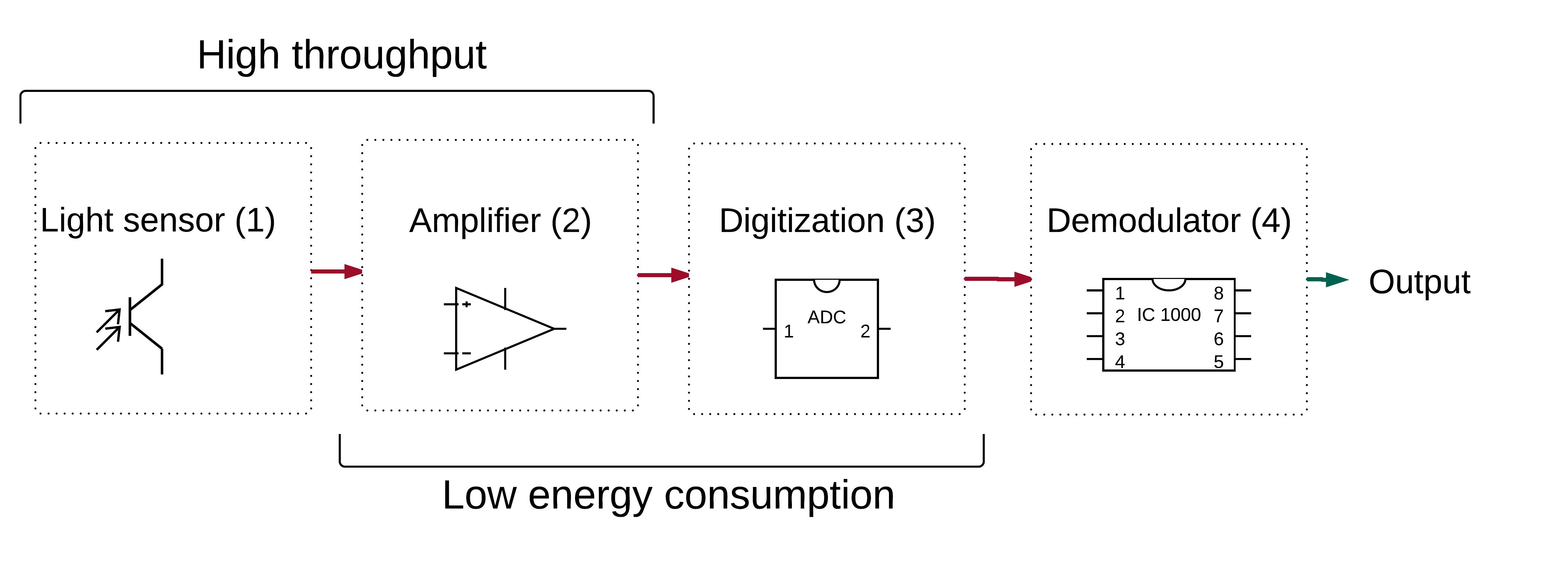}
  \caption{Abstract architecture for a VLC receiver. \capt{Different stages are responsible for optimizing different performance metrics. The architecture provides guidance on where to intervene to attain given performance goals.} }
  \label{fig:receiverarch}
\end{figure}

\section{Design}
\label{design}

The fundamental operation in a VLC receiver is to track and digitize changes in light intensity levels generated by a light transmitter, such as controllable LEDs or bulbs.
At the receiver, the light intensity changes are eventually translated into useful bits through a multi-stage pipeline~\cite{ledtoled}.
There is certainly not just one way to structure such a pipeline.
Moreover, in the general case it is unclear what stage in the pipeline would be mainly responsible for addressing a specific requirement, such as energy consumption or throughput. 

We structure our work around the abstract architecture of Figure~\ref{fig:receiverarch}.
The architecture helps us separate different functional stages in the pipeline, define their in\-put/out\-put relations, and focus the selection of individual components based on specific requirements.
In fact, we recognize that specific stages in the pipeline are mainly responsible for certain performance metrics.

Matching the architecture of Figure~\ref{fig:receiverarch} against current state of the art~\cite{epsilon,ledtoled,darkVLC,openvlchotwireless,zhang2015dancing} and commercial\-ly available components leads us to the following insights:
\begin{enumerate}
  \item At the first stage of the pipeline, some form of light sensor must convert the incident light to electric energy.
    We recognize that when employed in VLC receivers, the sensor's ability to rapidly track changes in light levels \emph{impacts the throughput}~\cite{openvlchotwireless}.
  \item Often, the sensor output must be amplified to boost the energy levels to useful degrees.
    Such a stage is responsible for both the \emph{throughput performance}, in that it must match the dynamics of the sensor output, and the \emph{energy consumption}, as amplifiers may incur significant energy costs~\cite{ltc6268}. 
\item Next, we must translate the signal from the analog to the digital domain.
This stage is typically most energy-hungry~\cite{Li:2015:REB:2699343.2699354,ambientbackscatter}, especially when using commonly available ADCs.
This reveals that an alternative design for this stage would be sought to \emph{reduce energy expenditures}.
\end{enumerate}

\begin{figure}[!tb]
\centering
  \includegraphics[width=0.35\textwidth]{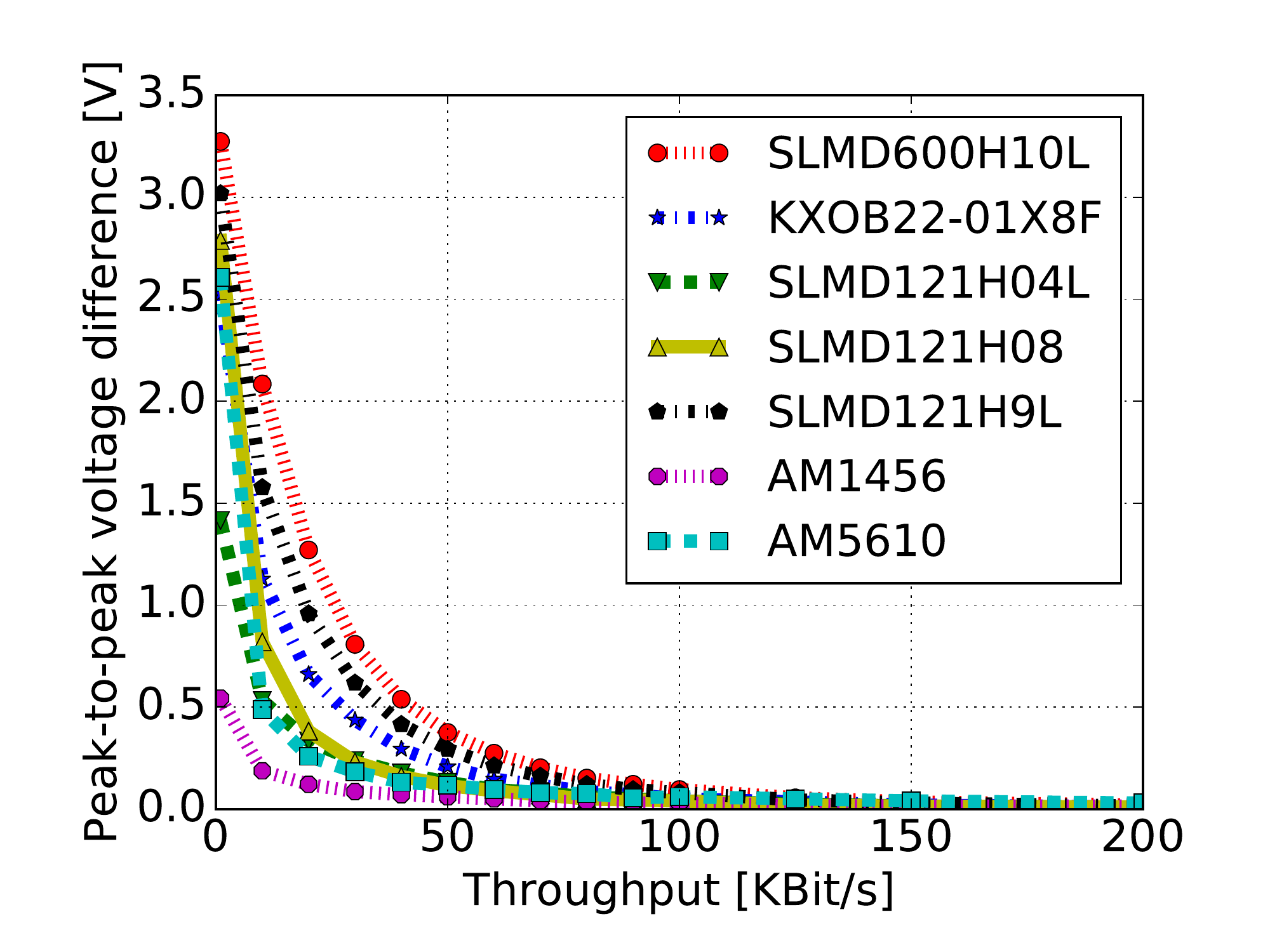}

        \caption{Performance of commercially-available solar cells. \capt{Distinguishing between the ON and OFF states of a VLC transmitter becomes difficult at higher bitrates since the peak-to-peak voltage difference becomes too small.} }
  \label{fig:solarcellbitrate}

\end{figure}

We show next how these insights leads to two different designs, each optimized for a different performance metric relevant in the wearable domain.
We note, the designs we present are instantiated using off-the-shelf components. ASIC versions of the designs can further reduce the size of the receivers.

\subsection{Ultra-Low-Power Receiver}
\label{ulp}

\begin{figure}[!tb]
\subfigure[Schematic.]{\label{fig:solarschematic}\includegraphics[width=\linewidth]{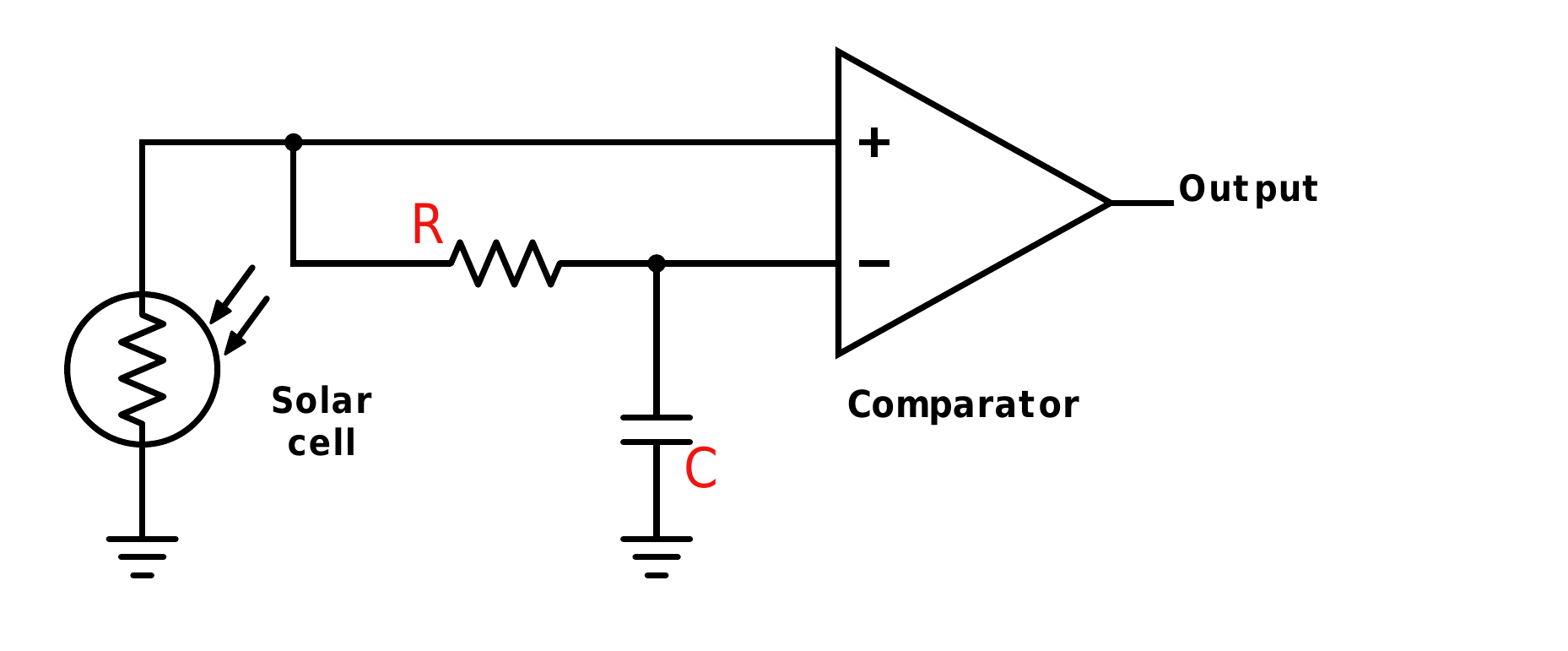}}
\subfigure[Energy-harvesting WISP device interfaced to a \ulp receiver.]{\label{fig:wispulp}\includegraphics[width=1.0\linewidth]{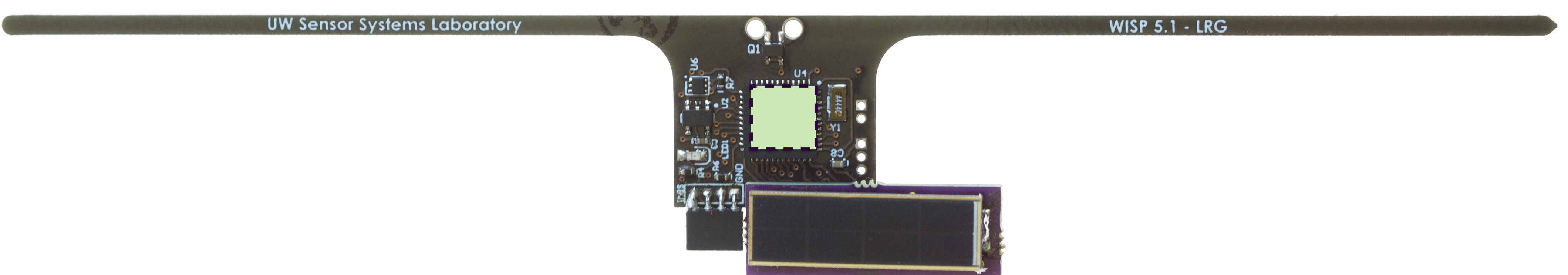}}
\caption{\ulp receiver. \capt{A solar cell replaces the light sensor and amplification stages. This enables energy savings at the expenses of maximum achievable throughput.}}
\label{fig:ulpreceiver}
\end{figure}

\begin{figure*}[!tb]
\centering
  \includegraphics[width=\textwidth]{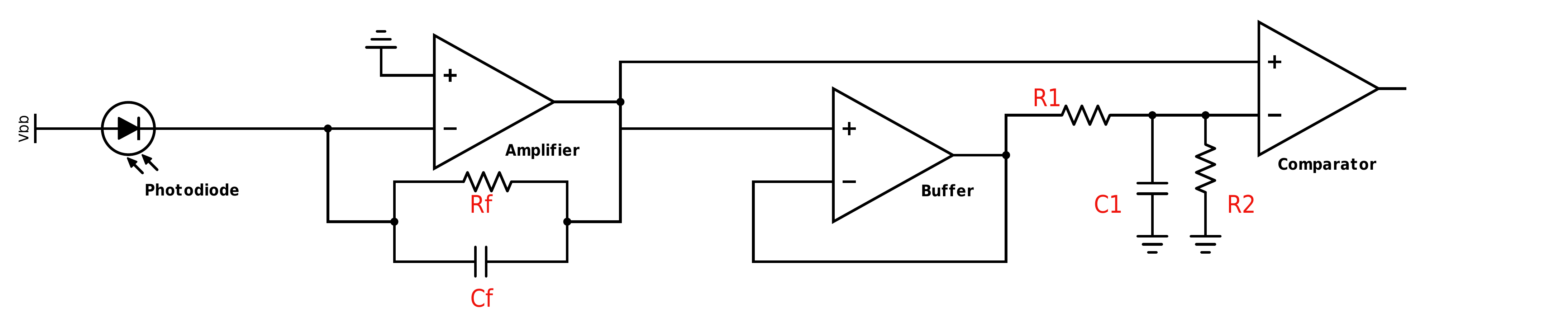}
  \caption{\highspeed receiver schematic. \capt{The photodiode and transimpedance amplifier allow the device to obtain high throughput. Thresholding circuit  digitises signal, and also helps to support mobility.} }
  \label{fig:hsschematics}
\end{figure*}

To optimize energy consumption in applications employing extremely resource-constrained wearables, we explore a new design both at the amplification and at the analog-to-digital conversion stage. 
Figure~\ref{fig:ulpreceiver} shows our \ulp prototype.
Key design choices are the use of a solar cell for light sensing to spare the amplification stages, and of a thresholding circuit in place of a standard ADC to digitize the signal.

\fakepar{Solar cells}  A solar cell is similar to a photodiode in construction; the voltage across the terminals changes with light fluctuations, making it applicable as a light sensor.
Solar cells, however, are optimized for a very different purpose compared to photodiodes, that is, to collect as much incoming energy as possible.
As a result, solar cells aim at generating the highest possible voltage at the output.
To reduce energy consumption, we turn this feature to our advantage by sparing the amplification stage.

We therefore employ a solar cell to implement the first two stages in Figure~\ref{fig:receiverarch}.
This, however, does not come without disadvantages.
Solar cells exhibit capacitance effects, which affect their ability to respond quickly to changing light conditions.
The pace of such changes must match the bitrate generated by the VLC transmitter.
As a result, solar cells likely limit the achievable throughput, and the selection of the component becomes crucial.

\fakepar{Choosing solar cells} We evaluate the responsiveness to changes in light levels of several existing solar cells.
As transmitter device, we use a controllable LED connected to a pulse-wave generator that creates an alternating sequence of 1s (LED on) and 0s (LED off).
We test seven solar cells with a form factor suitable for wearable applications. 
Five of these are monocrystalline with different dimensions and parameters, while two are amorphous silicon cells.
We connect all solar cells to the ADC of a logic analyzer to find the peak-to-peak difference in the signal amplitude.


Figure~\ref{fig:solarcellbitrate} demonstrates that all seven solar cells exhibit similar behaviors.
As the sending bitrate increases, the solar cells' ability to distinguish between the two LED states diminishes. 
For our \ulp receiver, we choose the SLMD121H04L~\cite{SLMD121H04L}~(< \$6) cell. 
While showing a performance similar to the other cells in terms of responsiveness in the visible spectrum, it is also small and generates high short-circuit currents. 
This opens up the possibility of harvesting energy for the host device directly from the VLC receiver.

\fakepar{Thresholding} At the third stage in the pipeline of Figure~\ref{fig:receiverarch}, the analog signal is to be converted to the digital domain.
Common ADCs can perform such conversion.
However, their operation is extremely costly in terms of energy consumption~\cite{Li:2015:REB:2699343.2699354,ambientbackscatter}.

To address this issue, we employ a thresholding circuit in place of a common ADC at this stage.
The thresholding circuit converts the voltage fluctuations output by the solar cells to binary values.
To that end, the circuit dynamically calculates a moving average of the running signal and compares it to the output of the solar cell to digitize the signal.
The circuit therefore autonomously adapts to fluctuating light conditions though changes in the moving average.
This turns out to be an asset because of the light dynamics induced by mobility, as mentioned in the Introduction.
In traditional designs, such a functionality can only be emulated in software, at the expense of frequent energy-hungry ADC operations.
Owing to its low energy consumption, we integrate a TS881~\cite{ts881} comparator~( < \$1) from ST Microelectronics.

To the best of our knowledge, our \ulp receiver is the first to exploit a solar cell coupled to a thresholding circuit to achieve a power consumption as low as \SI{0.5}{\micro\watt}, as we demonstrate in Section~\ref{sec:eval}.
On the other hand, the use of a thresholding circuit instead of a common ADC restricts the choice of modulation scheme, in that the circuit only provides binary outputs.
This makes it difficult to support complex modulation schemes like OFDM or PAM.
It does not prevent, however, supporting schemes most commonly employed in VLC such as FSK, OOK, PPM and PWM~\cite{6163585}.


\subsection{High-Speed Receiver} 
\label{sec:hs}

\begin{figure}[!tb]
\centering
\subfigure{\includegraphics[width=0.75\linewidth]{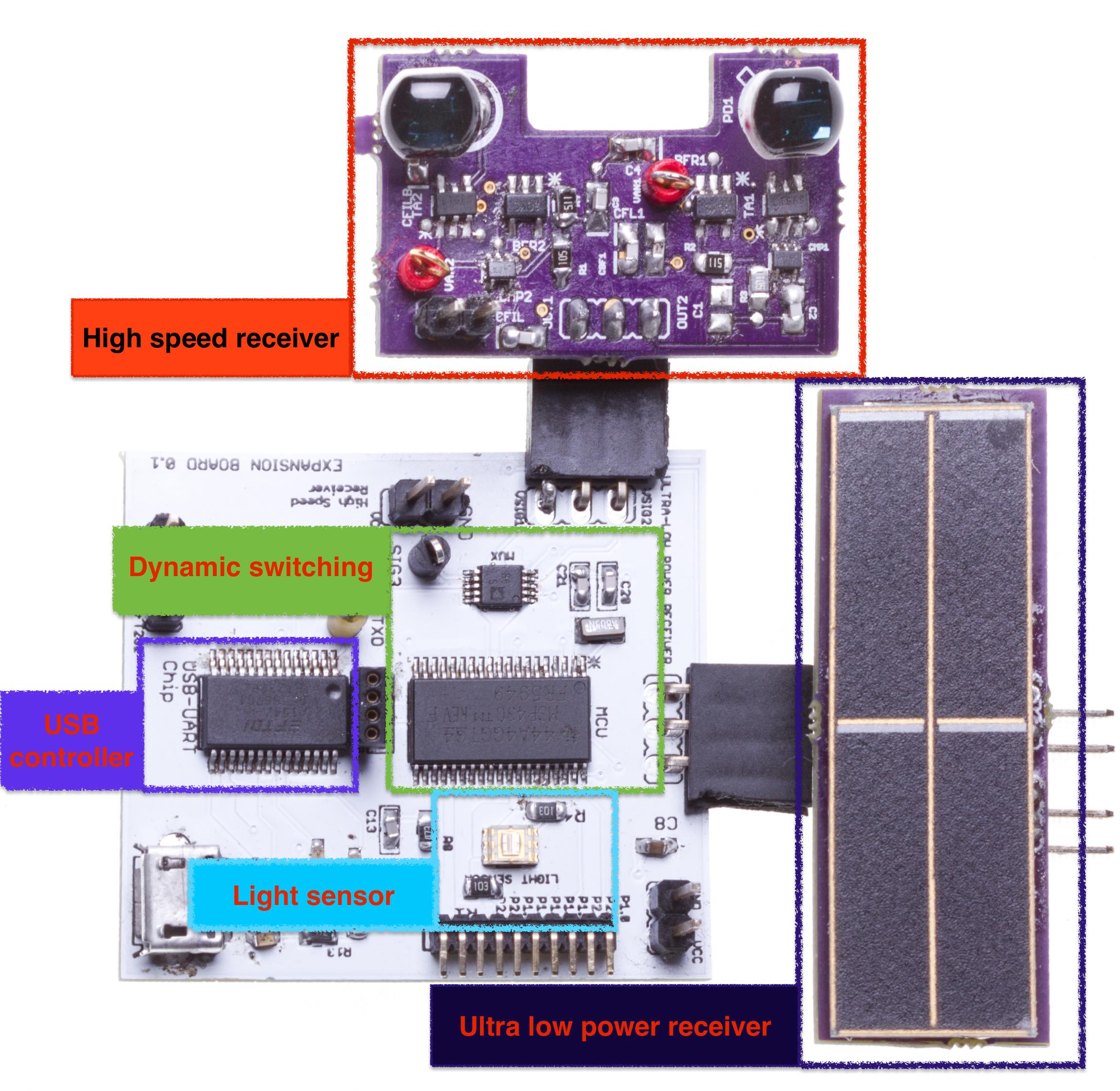}}
\caption{\platform unit prototype. \capt{ Combing three different receivers  helps to tackle rapid changes in light intensity due to mobility.} }
\label{figure:Integraion}
\end{figure}

To fulfill the high throughput requirement, according to Figure~\ref{fig:receiverarch} we focus on the sensing and amplification stages.
Key design choices are the selection of a highly sensitive photodiode and a transimpedance amplifier (TIA) able to match its dynamics as well as the use of a thresholding circuit.
Figure~\ref{fig:hsschematics} shows the schematic of our \highspeed receiver, whereas Figure~\ref{figure:Integraion} depicts the corresponding prototype.

\fakepar{Light sensor and amplifier} We use a SLD-70BG~\cite{sld70bg} photodiode~( < \$7) from Silonex.
This particular photodiode is most sensitive at a wavelength of to \SI{550}{\nano\meter}, which falls in the visible light spectrum.
This allows the sensor to react rapidly to changes in light intensity in the most commonly expected operational conditions.

We use a transimpedance amplifier (TIA) at the next stage.
A TIA converts small currents to a corresponding amplified voltage.
Crucial to its operation is the gain configuration.
This must be carefully set because a TIA can only generate a certain maximum voltage for a fixed supply.
If the gain is too high and the output signal reaches the maximum level, the TIA is said to be \emph{saturated}.
At that point, changes in light intensity level go undetected at the output of the TIA.
If the gain is too low, the signal to noise ratio (SNR) at the TIA output might be too low to differ between light intensity levels.
In addition, TIAs have a fixed gain bandwidth product (GBP), that is, if the gain is increased, the bandwidth decreases and vice-versa.

We exemplify these issues with an experiment.
We probe the receiver using a controlled LED connected to a waveform generator~(AWG).
We change the gain of the receiver by altering the resistor value $R_f$ of the TIA.
For a given gain setting, we measure the minimum light levels that ensure the receiver is able
to receive transmissions, along with the maximum light levels at which the receiver can continue receiving transmission before the TIA starts to get saturated.

 \begin{figure}[!tb]
\centering
\mbox{
    \subfigure[Minimum light levels to ensure conversion to digital.\label{minimum:gain}]{\includegraphics[width=0.47\linewidth]{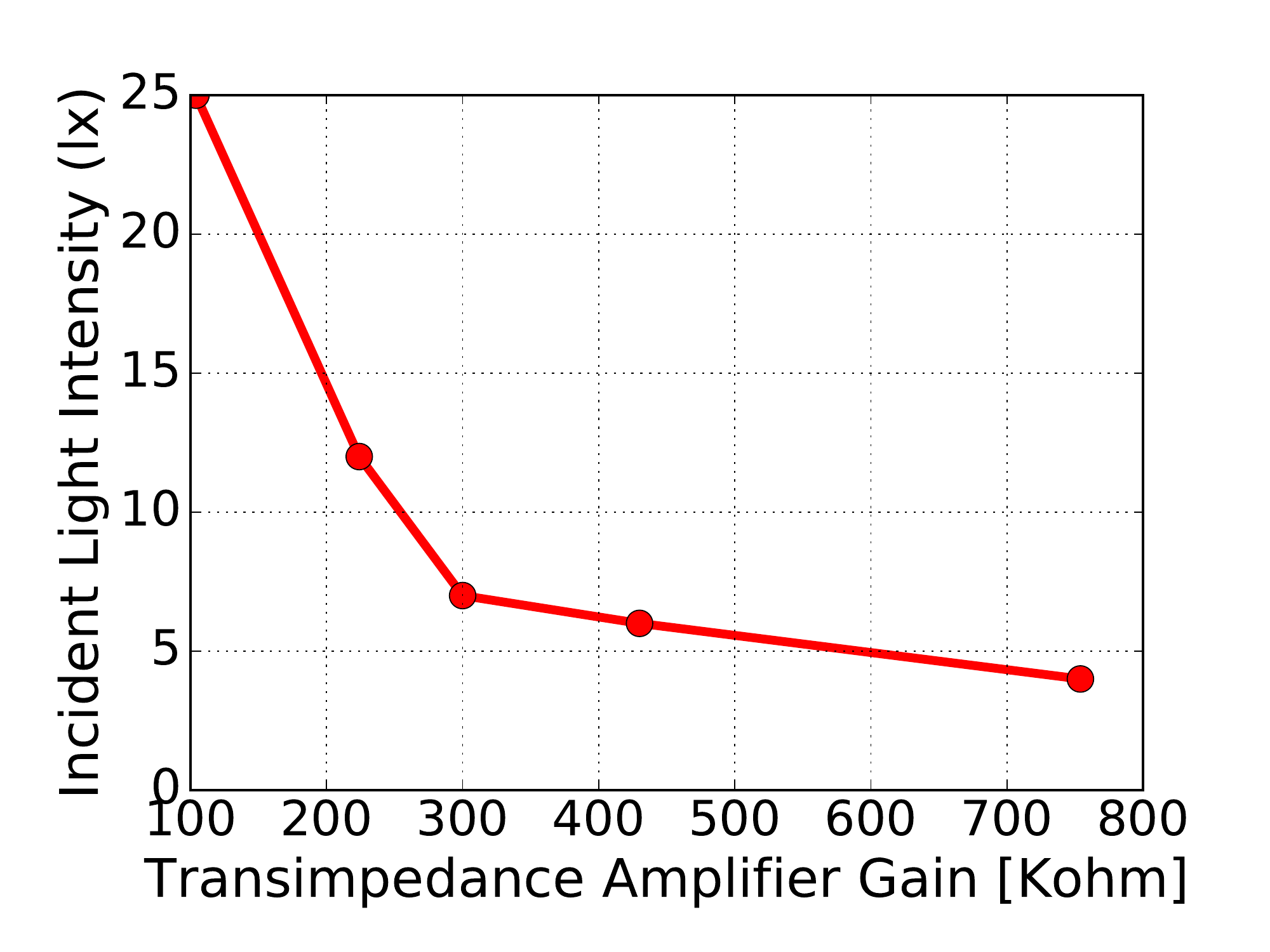}}
}
    \subfigure[Maximum light levels before saturation.\label{maximum:gain}]{\includegraphics[width=0.47\linewidth]{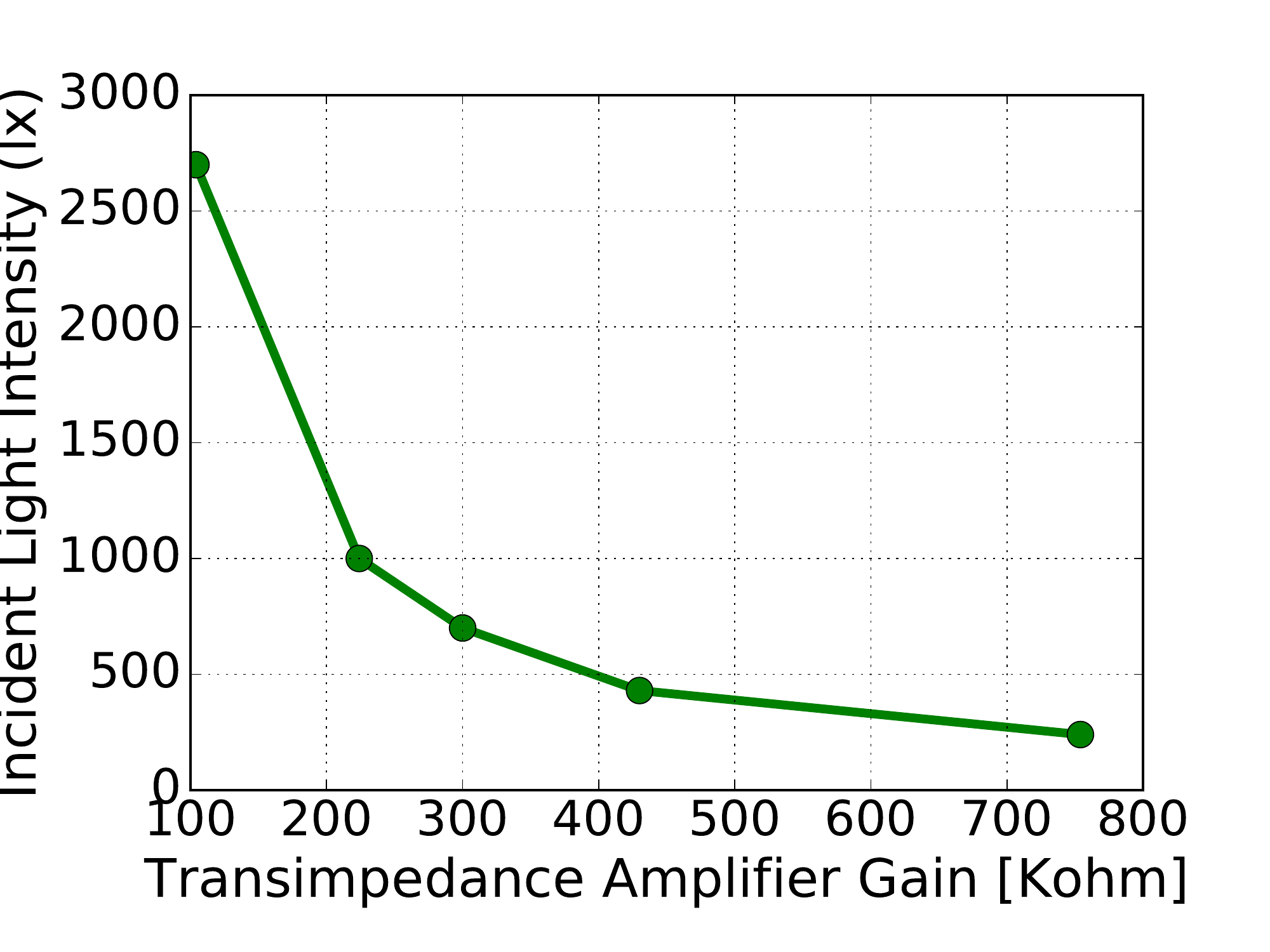}}
\caption{Amplifier gain compared to light levels. \capt{Identifying a gain setting that works under diverse light conditions is difficult. A high gain causes saturation even under moderate light levels. A low gain causes the analog to digital conversion to fail at low light levels.}}
\label{fig:gainvslight}
\end{figure}

Figure~\ref{fig:gainvslight} shows the results.
As the gain is increased, the minimum light levels needed to operate decreases.
For example, Figure~\ref{minimum:gain} indicates that when the resistor controlling the
gain is set to \SI{756}{\kilo\ohm}, the receiver can operate at even extremely low
light levels of \SI{4}{\lux}.  
The same gain setting, however, saturates the TIA easily; Figure~\ref{maximum:gain} shows 
this happening below \SI{240}{\lux}, corresponding to moderate levels of natural lighting.
Similarly, at a low gain setting of \SI{104}{\kilo\ohm}, the TIA does not saturate 
even under very bright light of \mbox{$\approx$ \SI{2800}{\lux}}.
The receiver, however, fails to operate at light levels lower than \SI{25}{\lux}.
These extremes are, however, representative of the wildly variable light conditions one may experience under mobility, as shown in Figure~\ref{fig:motivation}. 

To enable operation under diverse light conditions, we design the \highspeed receiver
with a dual photodiode configuration, thus providing the ability to set  two gain configurations.
The dual photodiode enables 
switching between the two configurations as a function of the prevailing light conditions.
We call these configurations \hgain and \lgain receiver, respectively.
We configure the high gain setting on the photodiode to be \SI{754}{\kilo\ohm}, and the low gain configuration to be \SI{104}{\kilo\ohm}.
Note, however, that increasing the gain comes at the cost of reduced bandwidth because of the fixed bandwidth-gain product we mentioned earlier.
As a TIA, we use the Linear Technology LTC6268~\cite{ltc6268}~(< \$7) because of high GBP and low power consumption.

\fakepar{Thresholding} Similar to the \ulp receiver, here again we use a thresholding circuit to convert the signal to the digital domain.
We can thus enjoy the same benefits as in the \ulp with the rapidly varying light conditions induced by mobility.
To support the high throughput that the early stages in this design would yield, we couple a comparator circuit with a low-pass filter, as shown in Figure~\ref{fig:hsschematics}.
The latter computes the running average of the signal; the comparator mat\-ches the averaged signal with the input.

We use a ON Semiconductor NCS2200~\cite{ncs2200} as the comparator~( < \$1), owing to its low propagation time and energy consumption. We note that similar designs of thresholding circuits are also used 
elsewhere, for example, ambient backscatter receivers~\cite{turbobackscatter,wifibackscatter,ambientbackscatter,HitchHike}, visible-light backscatter~\cite{Li:2015:REB:2699343.2699354}, and for ultra low-power gesture detection~\cite{allsee}.

%% file: platform.tex
\section{Dynamic Switching}
\label{sec:platform}

  	



The three VLC receivers described in Section~\ref{design} provide efficient performance only in a subset of the possible light conditions that wearable devices may experience.
To extend their functioning across the board, we design a means  to dynamically switch between the three receivers based on current light conditions and as a function of performance goals.

\fakepar{Switching logic} We consider two key performance goals: throughput and energy consumption.
The logic to switch between receivers is essentially the same in the two cases; it only requires minor adaptations as a function of the performance goal.

When optimizing for throughput, we probe the operating light conditions at the maximum possible rate.
Whenever the switching logic executes directly on a sufficiently powerful wearable device, such a sampling may make use of any light sensor already available there.
When running on the dedicated integration unit we describe next, this happens through a standard photodiode via ADC calls.
In either case, the rapid sampling ensures that switching to the appropriate receiver occurs as soon as possible.
To identify the appropriate receiver, we match the operating ranges of the individual receivers against current light conditions.
In case multiple options are available, we select the receiver with the maximum supported throughput.

The switching logic when optimizing for energy is similar.
The main difference is the sampling of light conditions, which would be prohibitively expensive in terms of energy consumption if performed at high frequency.
To tame this issue, we sample the operating light conditions every $\Delta$ time units.
In this case, should multiple options match the operating ranges, we favor the receiver with better energy efficiency.

\fakepar{Integration unit} The wearable device may not be able to host the switching logic; either because of resource scarcity or due to hardware integration issues.

To address this issue, we design a custom integration unit that off-loads such processing from the wearable device. 
The unit includes a low-power MCU dedicated only to manage the dynamic switching between the three VLC receivers as light conditions change, and tasked with the necessary demodulation logic.
As a result, the integration unit presents the three VLC receivers as a single unit to the wearable device, masking the underlying complexity which eases the integration.

Our prototype, shown in Figure~\ref{figure:Integraion}, mounts a Texas Instruments MSP430FR5949~\cite{msp430fr5949} (< \$6) chip owing to various low-power modes available.
It features 64 $KB$ of FRAM,  2 $KB$ of SRAM, and a maximum running frequency at \SI{24}{\mega\hertz}.
A TSL2561~\cite{TSL2561} light sensor (< \$3) serves to measure the current light conditions, providing the necessary input to the switching logic.

We design the switching circuitry to enable both high speed operations and low power consumption. 
To control the two \highspeed receivers, we use an ultra low power switch TI TPS22944~\cite{tps22944} (< \$1).
This particular device draws
only~$\approx$ \SI{1}{\micro\ampere} when quiescent, and can turn on within \SI{60}{\micro\second} to support the current draw of a receiver.
Because of the efficient energy performance of the \ulp receiver, we simply keep it always on.
To route the output from the three receivers, we use a four channel multiplexer chip AD ADG704~\cite{adg704} (< \$3), which draws less than $\approx$ \SI{0.001}{\micro\ampere} and enables switching between different receivers within \SI{20}{\nano\second}.

Finally, whenever the wearable device offers a USB interface, the integration unit can connect to it directly by means of a dedicated a UART-to-USB chip.
The chip demodulates the UART signal and further passes it serially through the USB.
Using such a chip as de-modulator is possible because the thresholding circuit used in all VLC receivers mimics the modulated digital signal.
Consequently, if an UART signal is transmitted, at the receiver we obtain a signal mirroring the transmitted UART signal.
We use an FTDI FT232~\cite{ft232} (< \$5) chip as USB controller.  



%% file: eval.tex
\section{Evaluation}
\label{sec:eval}

\begin{figure}[!tb]
\centering
\includegraphics[width=0.8\linewidth]{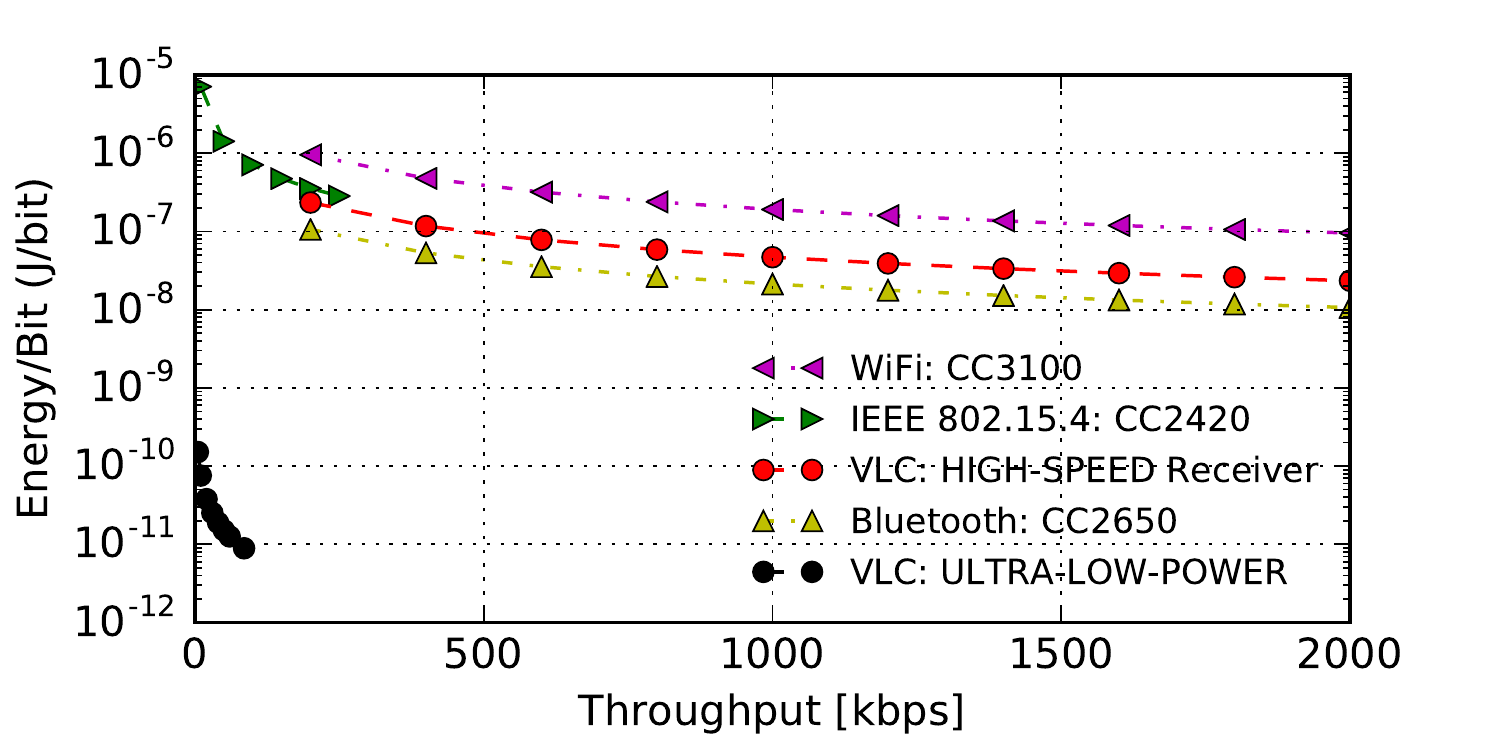}
\caption{Energy consumption/throughput comparison among VLC receivers and modern RF chips. \capt{The \ulp and \highspeed receivers distinctively meet their performance goals by retaining comparable performance in other metrics as existing RF chips or VLC receivers.}}
\label{fig:highspeedenergybit}
\end{figure}

We measure the performance of the prototypes we build in a range of different conditions.
Compared to RF transmissions, VLC is arguably less mature, especially when employed for wearable devices.
However, the performance of our prototypes turns out to be competitive with a range of modern RF chips, both in terms of throughput and energy consumption.
As a teaser for the results to come, Figure~\ref{fig:highspeedenergybit} depicts the performance of the \ulp and \highspeed receivers in the energy/through\-put plane, next to that of representative RF chips. 

We draw three fundamental observations from the experiments in this section:
\begin{enumerate}
  \item The \ulp receiver performs orders of magnitude better in energy consumption than state-of-the-art embedded VLC receivers and RF chips\, including those designed for low-power operation. Its thro\-ughput is comparable with state-of-the-art VLC receivers for embedded devices~\cite{openvlchotwireless}.
\item The \highspeed receiver, in either configuration, performs orders of magnitude better in throughput than state-of-the-art VLC receivers, and actually similarly to WiFi and Bluetooth chips in both throughput and energy consumption. 
\item In mobile settings, the switching logic rapidly adapts to fluctuating light conditions and switches to the best performing receiver; the integration unit provides high throughput through the USB interface. 
\end{enumerate}

In the following, we detail the settings used to obtain the results of Figure~\ref{fig:highspeedenergybit} as well as all those described in Section~\ref{sec:evalulp} to~\ref{sec:evalss}.

\subsection{Settings}

\fakepar{Setup}  We perform the experiments in our offices, where large windows allow natural light to enter and four fluorescent tube lights provide artificial illumination.
The latter flicker at the frequency of the AC signal (\SI{50}{\hertz}), which is much lower than the frequency used in VLC and hence bear no impact on the much faster changes due to data transmissions.
As VLC transmitter, we use an off-the-shelf LED rated to a maximum intensity of \SI{320}{\lumen} and operating at \SI{12}{\volt}.
We generate the input signal using a programmable waveform generator (PWG)~\cite{pwg}. 
We capture the output of the receiver using a logic analyzer~\cite{logicanalyzer}. We use LED driver circuit similar to the one used in our earlier work modBulb~\cite{modbulb}.

\begin{figure}[!tb]
\centering
\includegraphics[width=0.28\textwidth]{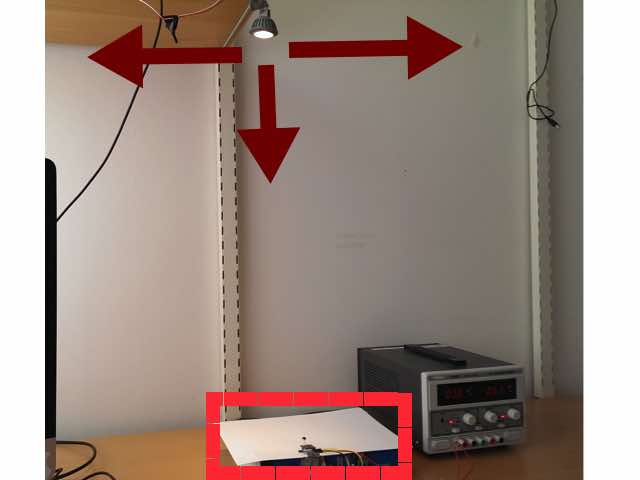}
\caption{Experiment setup. \capt{Arrows indicate the degrees of freedom for the transmitting LED that result in changes to the light intensity levels. The dashed box indicates the receiver and the co-located light sensor that measures incident light.}}
\label{expsetup}
\end{figure}

Figure~\ref{expsetup} illustrates the placement of VLC transmitter and receiver.
The LED is located about \SI{1.3}{\meter} away from the receiver.
We change the light intensity at the receiver by moving the position of the LED, as indicated by the red arrows.  
Vucic et al.\, in fact, show that the performance of a VLC receiver depends only on the light intensity levels and not on the length of the channel~\cite{5287385}.
To measure the ambient light conditions during the experiments, we use the unit $lux$, as is done in recent VLC systems~\cite{practicalhumansensing,darkVLC,passiveVLC}, and we measure the light conditions by co-locating the receiver with a TSL2561 light sensor~\cite{TSL2561}.


\fakepar{Metrics} We measure three key metrics, used extensively in low power communication systems~\cite{Bharadia:2015:BHT:2829988.2787490,wifibackscatter,ambientbackscatter}: (i) the \emph{energy per bit}, defined as 
the energy consumed to receive a single bit; (ii) the \emph{throughput}, defined as the number of useful bits received in a unit of time; and (iii) the \emph{bit error rate (BER)}, defined as number of bits received successfully as compared to those sent.



To determine the energy per bit, we connect a \SI{100}{\ohm} resistor in series with the receiver, and measure the potential drop using the logic analyzer.
To measure the BER, we send fixed-length packets using OOK as modulation scheme.
In a single round, we send 50 packets of \SI{256}{\byte}.
We repeat every experiment for three rounds, randomly regenerating the payload for packets sent in each round. We trace sent and received bits using the logic analyzer and measure the achievable throughput as the maximum transmission speed where the BER $\leq$ $10^{-3}$. 
We perform every experiment with three different levels of incident light, and in three different levels of ambient light.

\begin{table}[!tb]
\centering
\caption{Supported throughput at low light intensity levels for different VLC receivers}
\label{sensitvitytable}
\begin{tabular}{|l|l|l|l|}
\hline
\textbf{Light} & \textbf{High gain/} & \textbf{Low gain/} & \textbf{Ultra low}\\
\textbf{intensity} & \textbf{High speed} & \textbf{High speed} & \textbf{power} \\ 
\hline
\textless 4 lx           & 100 kbps                       & -                              & -                        \\ \hline
12 lx                    & 350 kbps                       & -                              & -                        \\ \hline
25 lx                    & 500 kbps                       & 100 kbps                       & 10kbps                   \\ \hline
50 lx                    & 600 kbps                       & 700 kbps                       & 30kbps                   \\ \hline
\end{tabular}
\end{table}


\begin{figure*}[!tb]
\centering
\hspace{-6mm}
\subfigure[Darkness\label{ulp:darkness}]{\includegraphics[width=0.33\linewidth]{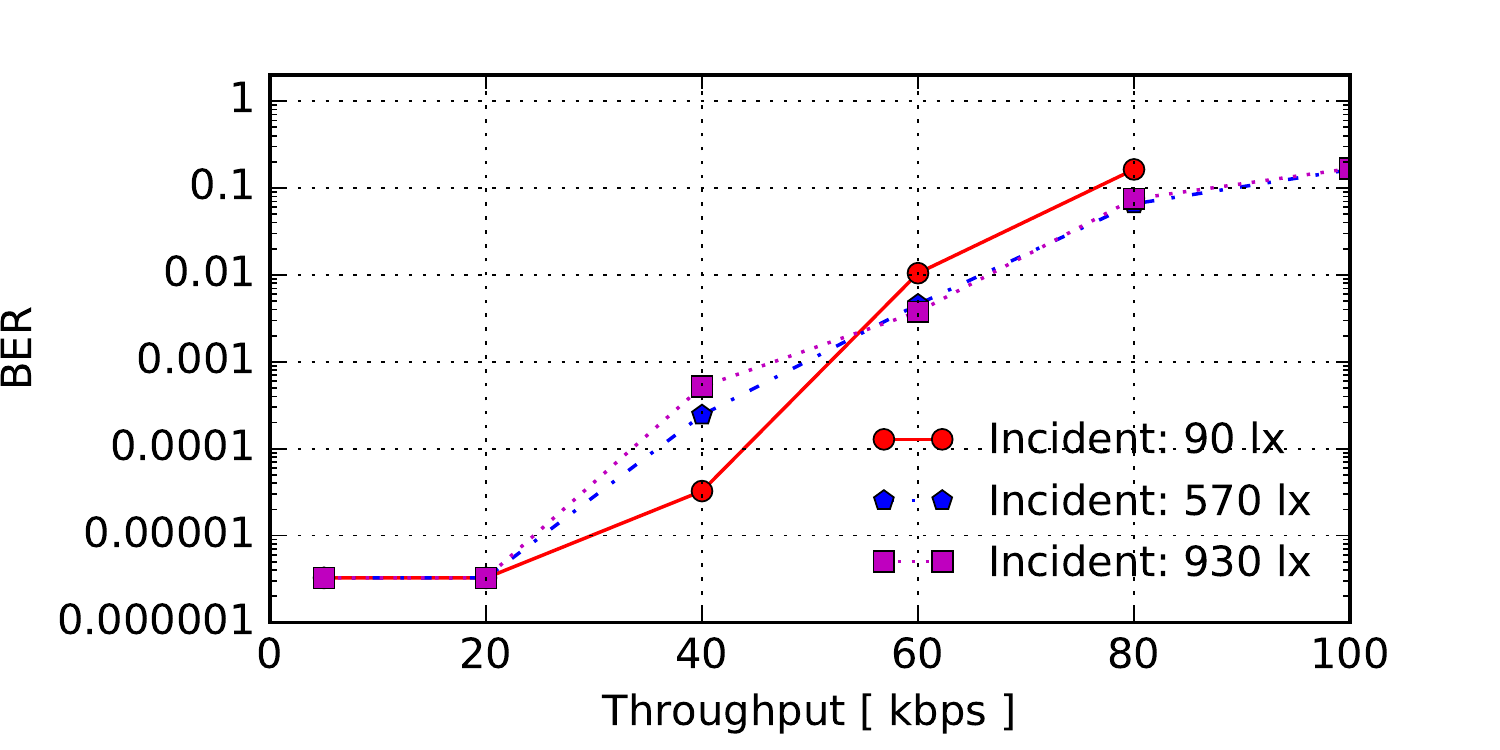}}\quad
\hspace{-6mm}
\subfigure[Indoor lighting (210 lx) \label{ulp:officelight}]{\includegraphics[width=0.33\linewidth]{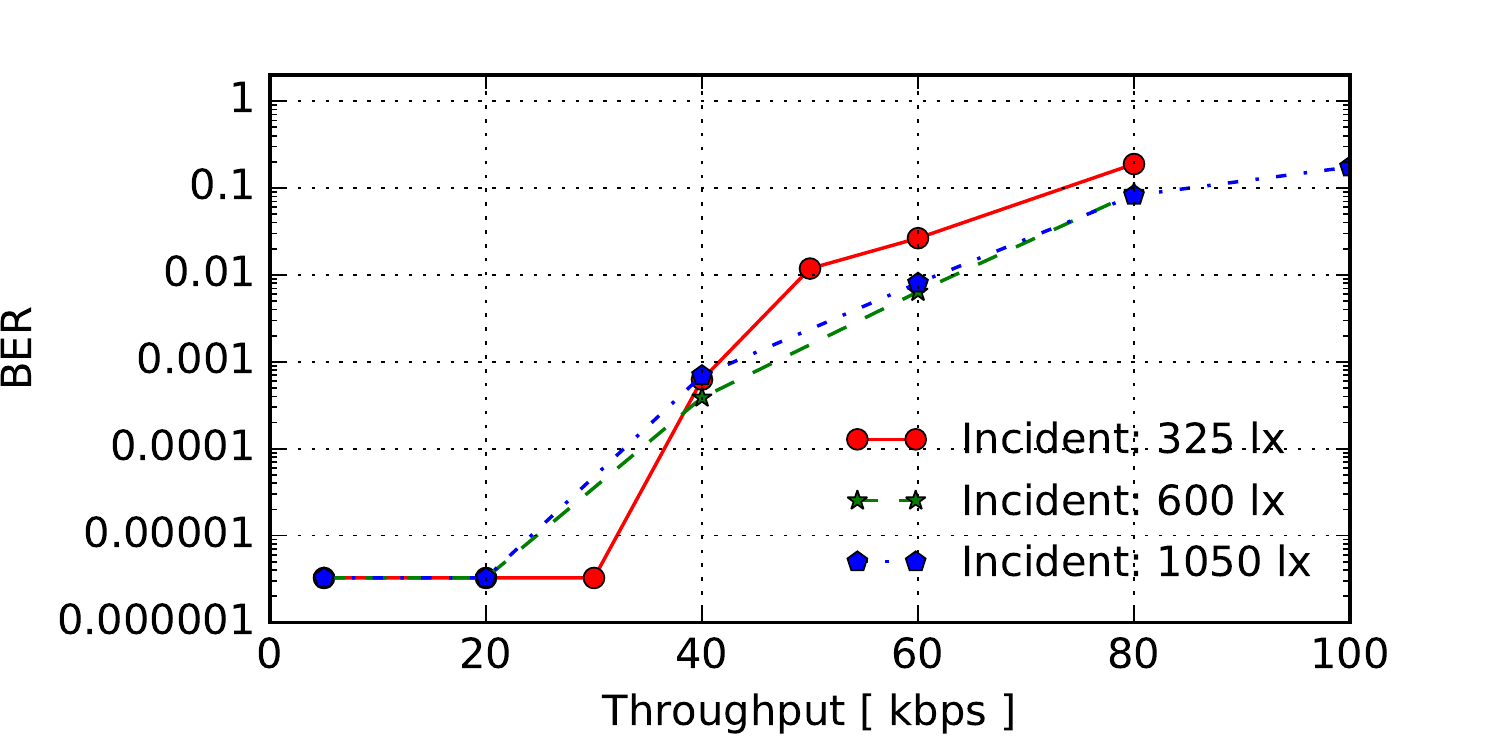}}\quad
\hspace{-6mm}
\subfigure[Natural lighting (350 lx)\label{ulp:natural}]{\includegraphics[width=0.33\linewidth]{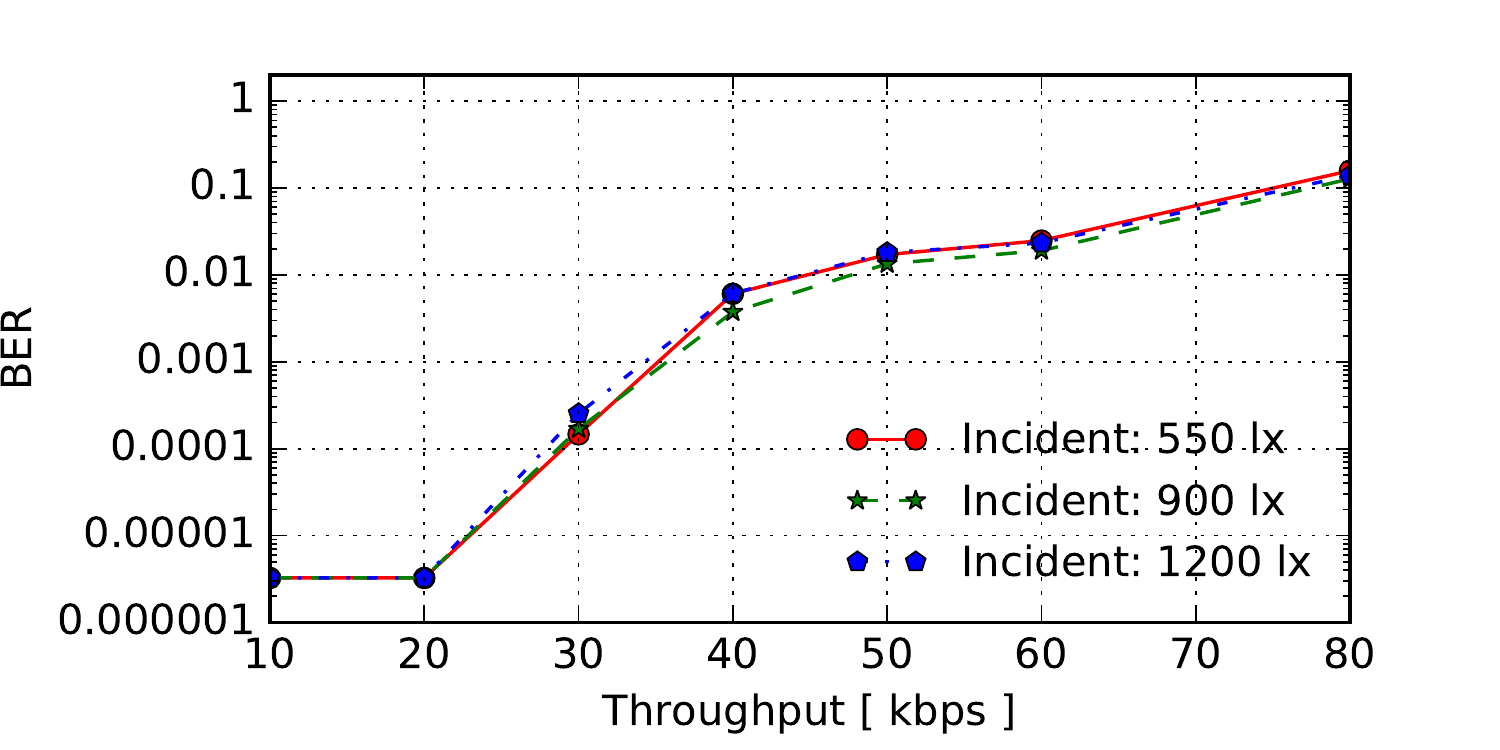}}\quad  
\vspace{-4mm}  
\caption{Throughput and BER of the \ulp receiver. \capt{The receiver performs well in diverse light conditions. As it is optimized for energy consumption, the receiver achieves a maximum throughput of 50 $kbps$ at a BER of $10^{-3}$.}}
\label{ulp:receivereval}
\end{figure*}

\subsection{\ulp Receiver} 
\label{sec:evalulp}


We design the \ulp receiver for energy-constrained wearable devices.
The \ulp establishes a specific design point that clearly reveals in the performance we measure.

\fakepar{Throughput} In this experiment, we measure the throughput achieved by the \ulp receiver in different ambient light conditions and incident light levels, as a function of the transmitter's data rate.
Figure~\ref{ulp:receivereval} depicts the results. 
Generally, as the transmitter's data rate increases, the BER starts to become significant. 
The reason for this behaviour is that the SNR decreases at higher data rates, which affects the receiver's ability to discern between the levels corresponding to 0 and 1 bits. Note the logarithmic scale for the BER in this and the following figures. 

The receiver tops at  60 $kbps$ at a BER of $10^{-3}$ in darkness, as shown in Figure~\ref{ulp:darkness}.
The maximum throughput slightly decreases as the levels of ambient lighting increase, as shown in Figure~\ref{ulp:officelight} and~\ref{ulp:natural}.
In these light conditions, the SNR decreases and therefore the BER starts to increase.
Interestingly, however,  under sufficiently bright light conditions ($\geq$ \SI{50}{\lux}) 
the intensity of incident light appears not to impact the maximum throughput,
as shown by the different curves in each of the charts of Figure~\ref{ulp:receivereval}.

\fakepar{Receiver sensitivity} We next investigate the receive sensitivity, that is, the minimum incident light levels required for the \ulp receiver to receive transmissions. In order to reduce power consumption, the \ulp receiver does not have an amplification stage which could reduce the receiver sensitivity.
In this experiment, the VLC transmitter varies its data rate as we change the orientation of the bulb to 
	find the minimum light levels required for successful operation of the receiver. Table~\ref{sensitvitytable} demonstrates the result of the experiment. 
	Our results show that the \ulp receiver operates at light levels corresponding to \SI{25}{\lux} at a throughput of 10 $kbps$. As we increase the incident light levels, the supported throughput improves to 30 $kbps$ corresponding to light levels of  \SI{50}{\lux} with little improvements thereafter. 
	 The results show that removing the amplification stage decreases the sensitivity, as compared to the \highspeed receiver.

\begin{figure*}[!tb]
\centering
\hspace{-6mm}
\subfigure[Mobility (10 kbps)\label{ulp:mobility}]{\includegraphics[width=0.33\linewidth]{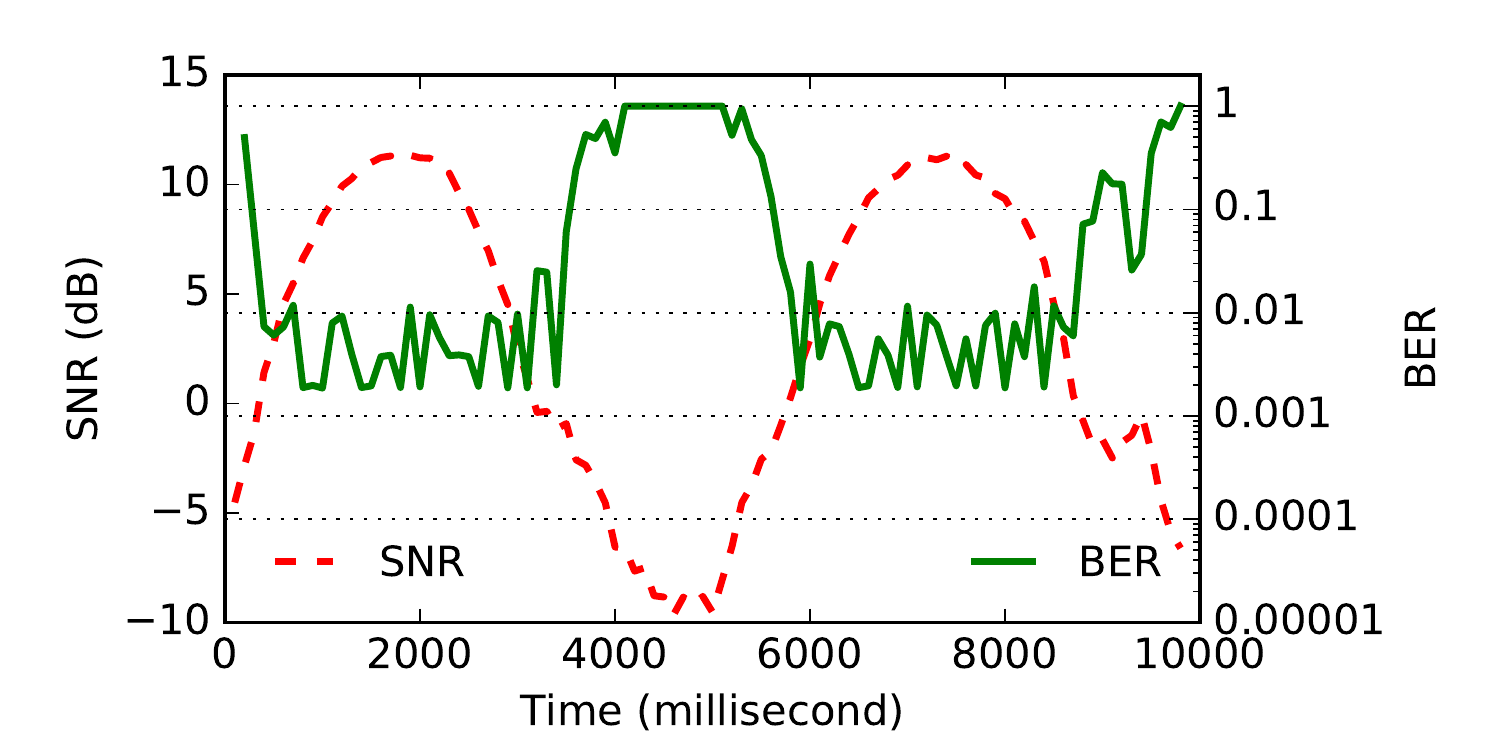}}\quad
\hspace{-6mm}
\subfigure[Orientation~(10 kbps)\label{ulp:orientation}]{\includegraphics[width=0.33\linewidth]{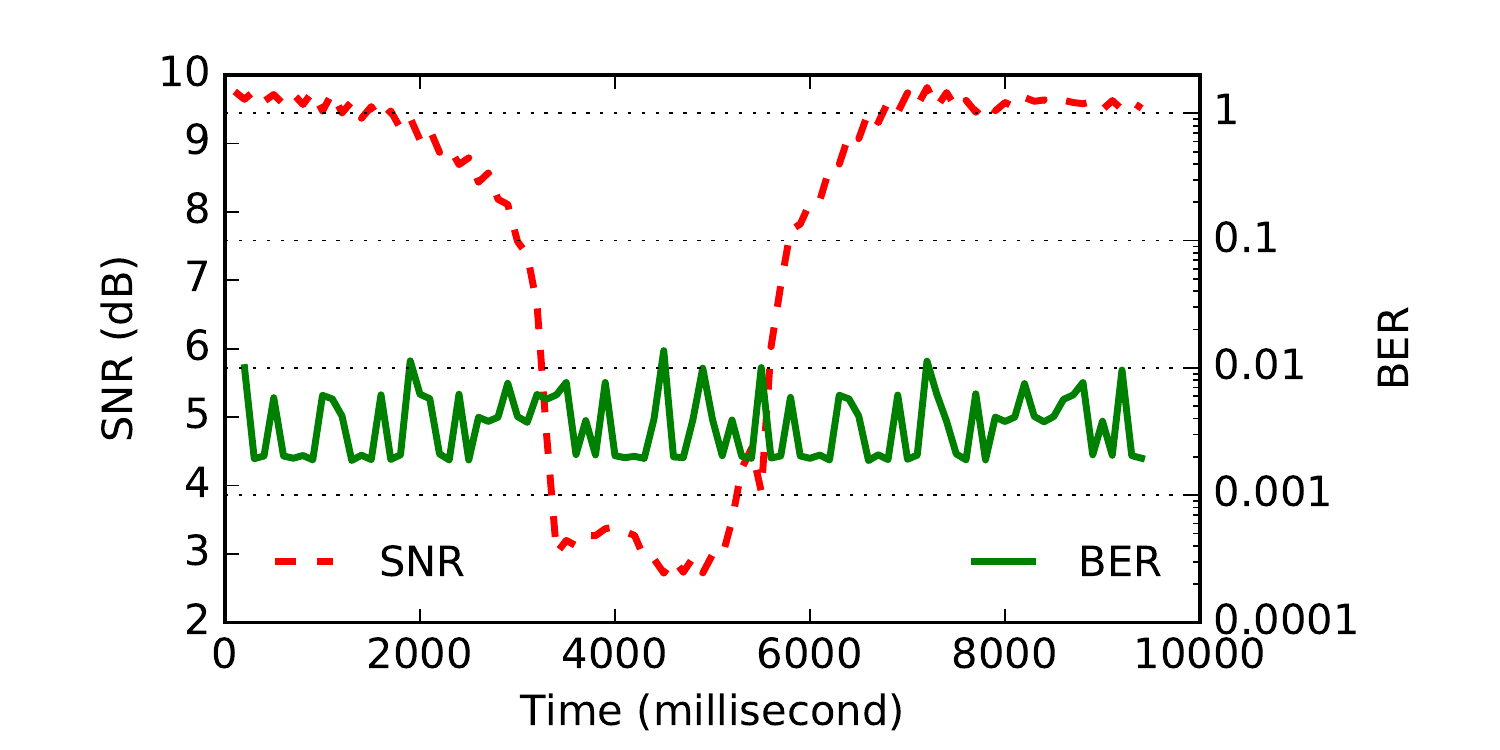}}\quad
\hspace{-6mm}
\subfigure[Orientation~(Flexible solar cell)\label{ulp:flexmobility}]{\includegraphics[width=0.33\linewidth]{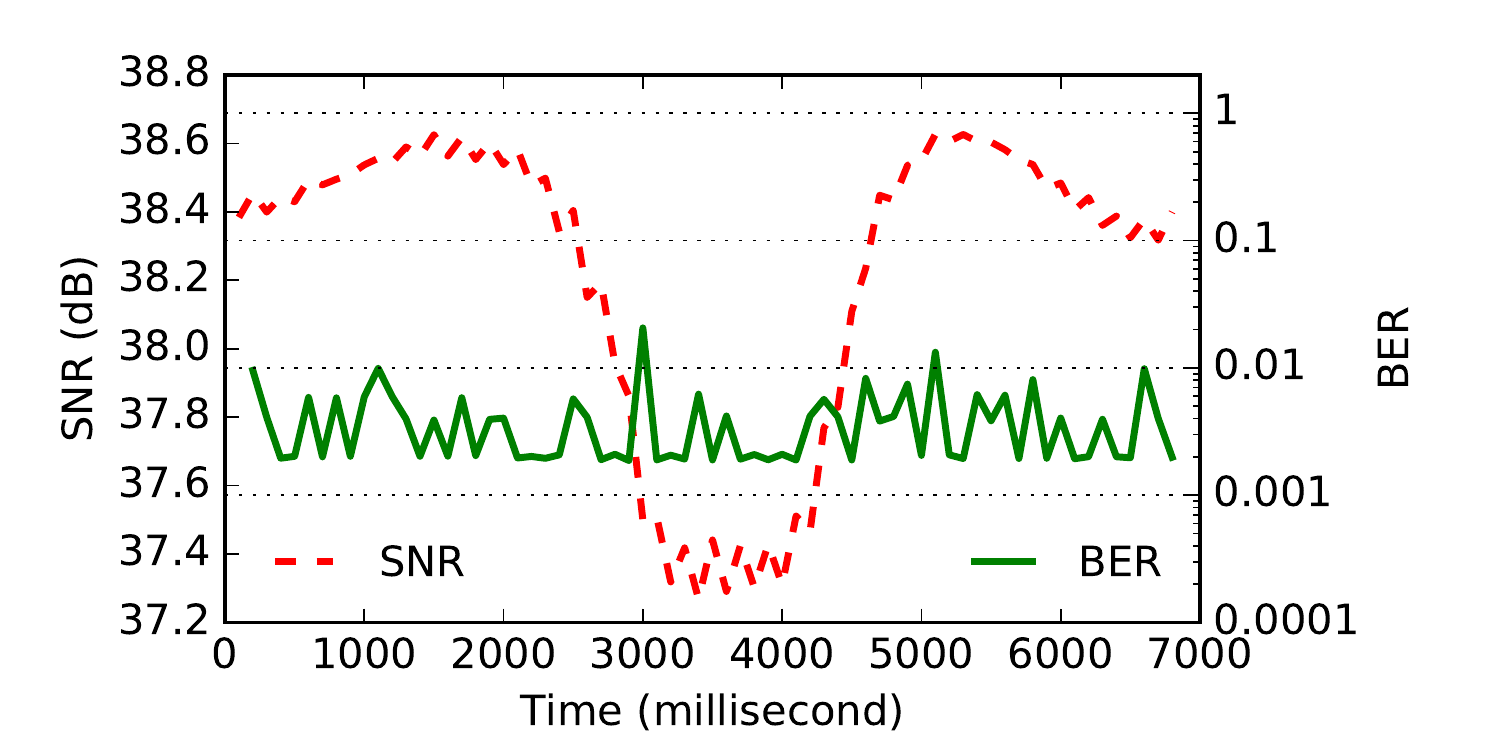}}
\vspace{-5mm}  
\caption{\ulp receiver in mobile scenarios. \capt{ The \ulp receiver achieves a low BER. When the SNR is positive, the thresholding circuit adapts to changing light conditions and interprets bits correctly. \ulp receivers performs well even when there is change in orientation.}}
\label{fig:ulp_mobility}
\end{figure*}

\begin{figure}[!tb]
\centering
\includegraphics[width=0.22\textwidth]{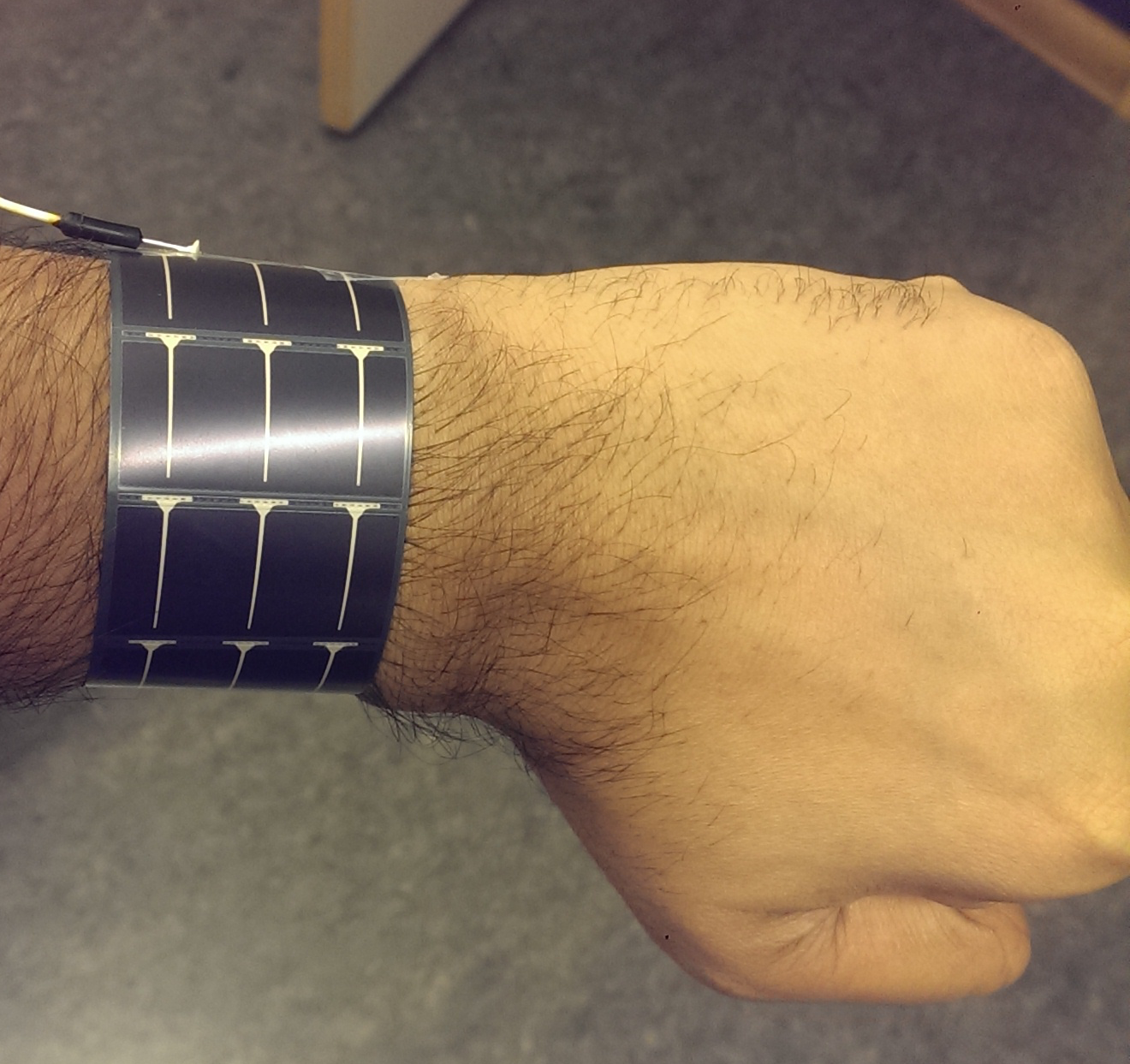}
\caption{\ulp receiver using flexible solar cell taped to wrist.}
\label{ulp:flexorientation}
\end{figure}

\begin{figure*}
\centering
\hspace{-6mm}
\subfigure[Darkness\label{receiver:darkness}]{\includegraphics[width=0.34\linewidth]{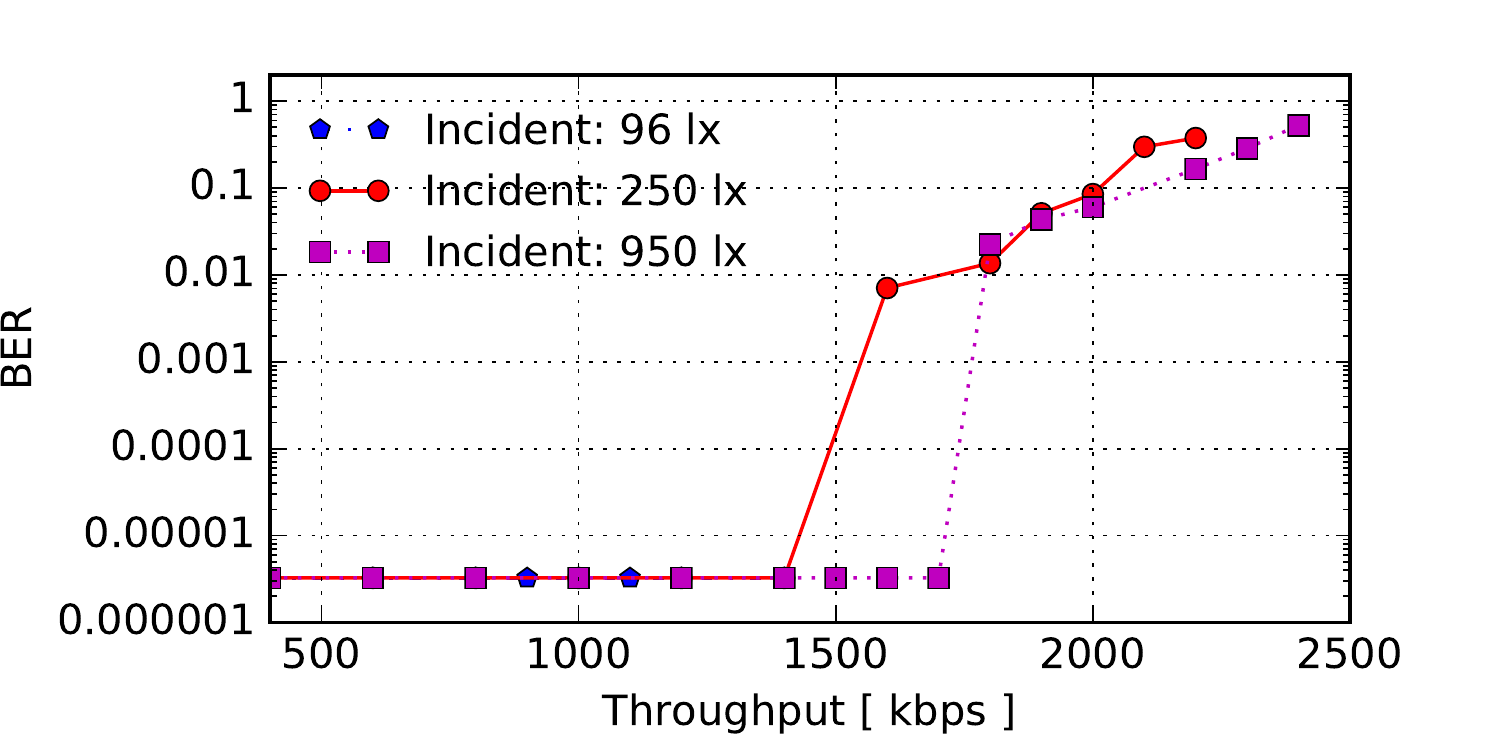}}\quad
\hspace{-6mm}
\subfigure[Indoor lighting (210 lx)\label{receiver:officelight}]{\includegraphics[width=0.34\linewidth]{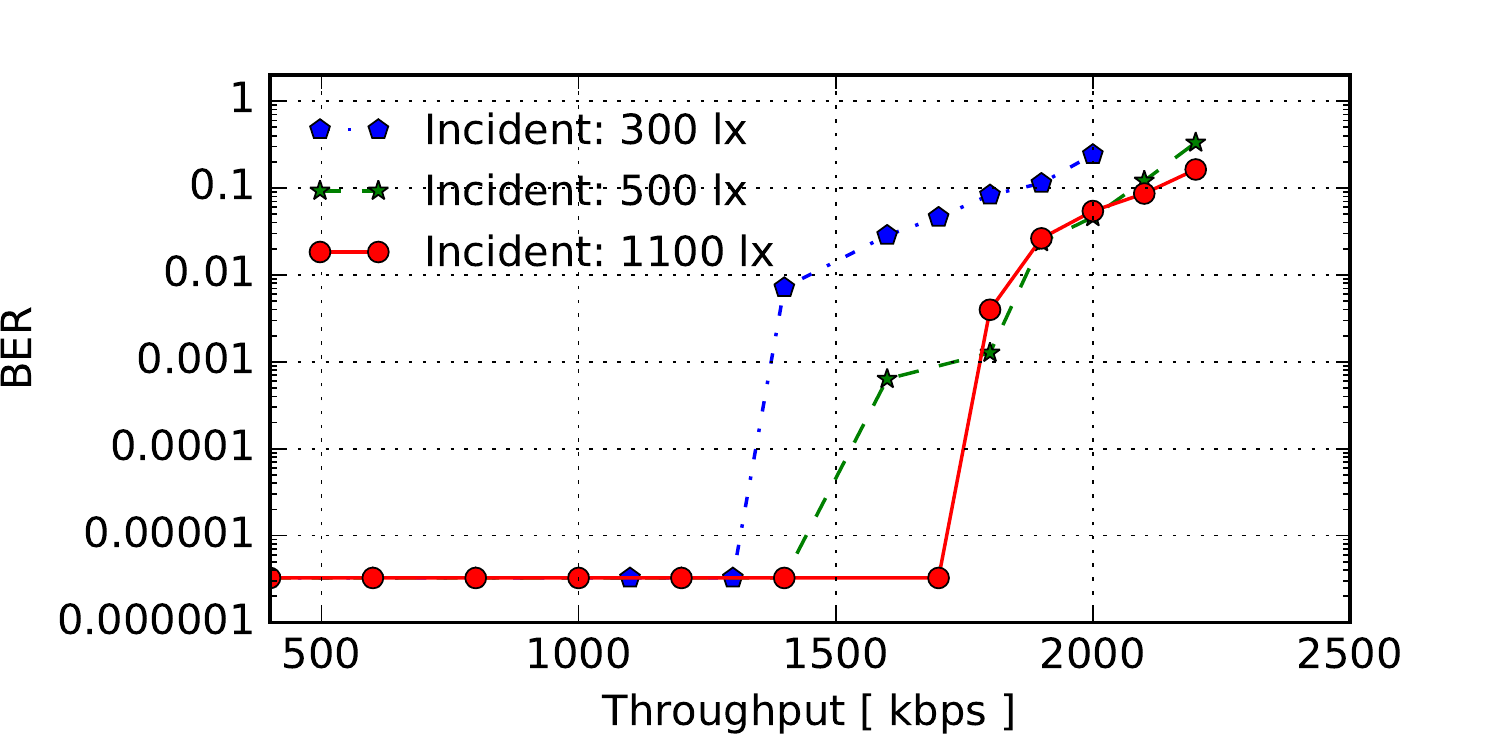}}\quad
\hspace{-6mm}
\subfigure[Natural lighting (350 lx)   \label{receiver:naturallight}]{\includegraphics[width=0.34\linewidth]{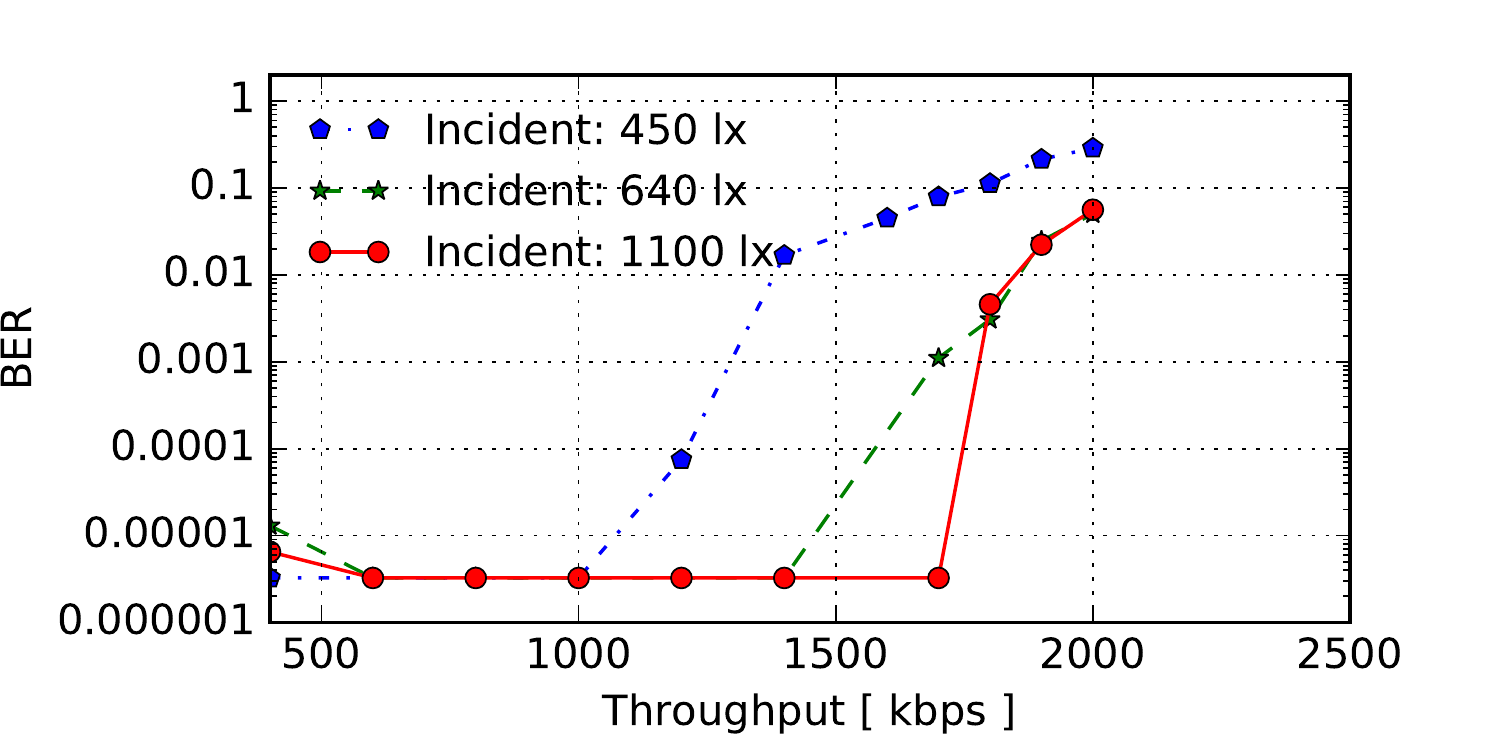}}\quad   
\vspace{-4mm} 
\caption{Throughput and BER of the \highspeed receiver. \capt{The receiver performs well in diverse light conditions. The supported bitrate increases with the incident light levels. We achieve a maximum throughput of 1700 $kbps$ at zero BER in darkness.}}
\label{receiver:highspeed}
\end{figure*}

\fakepar{Mobility} The 
\ulp receiver's thresholding circuit is able to track the variations in the incident light levels rapidly as it is implemented in hardware. This can help to, for example, keep pace with the fast changes in the incident light levels caused by mobility. Therefore, we investigate the performance of the receiver in mobile settings. We program the VLC transmitter to send a random sequence of bits at a data rate of 10 $kbps$. While transmitting, we change the incident light levels mimicking the changes observed under mobility as in Figure~\ref{fig:snr_motivation}.
	We capture both the analog and digital output of the \ulp receiver. We measure the SNR and BER every \SI{100}{\milli\second}. To calculate the SNR, we	take the signal observed at~\SI{25}{\lux} as the noise floor since this is the minimum light level the receiver can detect.

Figure~\ref{ulp:mobility} demonstrates that the SNR changes rapidly with the light intensity. In the figure, we see a lower change in the SNR as compared to Figure~\ref{fig:snr_motivation}, 
owing to the higher noise floor of the receiver.  
Figure~\ref{ulp:mobility} shows that the BER does not increase as the light intensity levels
change, except when the signal amplitude falls below the 
noise floor. The latter is indicated by the negative SNR in the graph. The results demonstrate that
dynamically changing light conditions
caused by mobility bear little impact on the achieved BER. This is because the thresholding circuit 
is able to track rapid changes in the SNR induced by mobility. 
At higher bitrates than 10 $kbps$, the BER is similar to the one shown in
Figure~\ref{ulp:darkness} even in mobile settings. We omit these results for brevity. 
 
 \fakepar{Orientation}  Changes in the orientation of a wearable device 
can also cause significant and rapid changes in the
SNR of the signal~\cite{zhang2015dancing}. We expect, that as in the mobile setting above, the receiver should be able to keep track of the transmitted bits as long as the SNR remains positive.  
To this end, we perform an experiment similar to the one above that
evaluated the receiver in a mobile setting.
Instead of changing the light intensity at the receiver, 
we rotate the receiver clockwise such that the angle between 
the VLC transmitter and receiver changes from \SI{0}{\degree} 
to \SI{90}{\degree} and back to \SI{0}{\degree}.
Figure~\ref{ulp:orientation} demonstrates
that the BER does not increase despite the
changes in the SNR due to the change in orientation.

We also instantiate a \ulp receiver using a flexible solar cell~\cite{powerfilm}. 
Flexible solar cells are appealing for wearable applications as they could
be pasted on clothing or even worn on the body to enable both communication 
and harvesting. As a proof of concept, we tape the flexible
solar cell based \ulp receiver on the wrist as shown in Figure~\ref{ulp:flexorientation}.
In an experiment similar to the one above, we slowly rotate our  wrist while
modulating the VLC transmitter at 10 $kbps$. Figure~\ref{ulp:flexmobility} demonstrates a very small drop of \SI{1}{\decibel} in the SNR and hence a low BER.

\begin{figure}[tb]
\centering
  \includegraphics[width=0.85\linewidth]{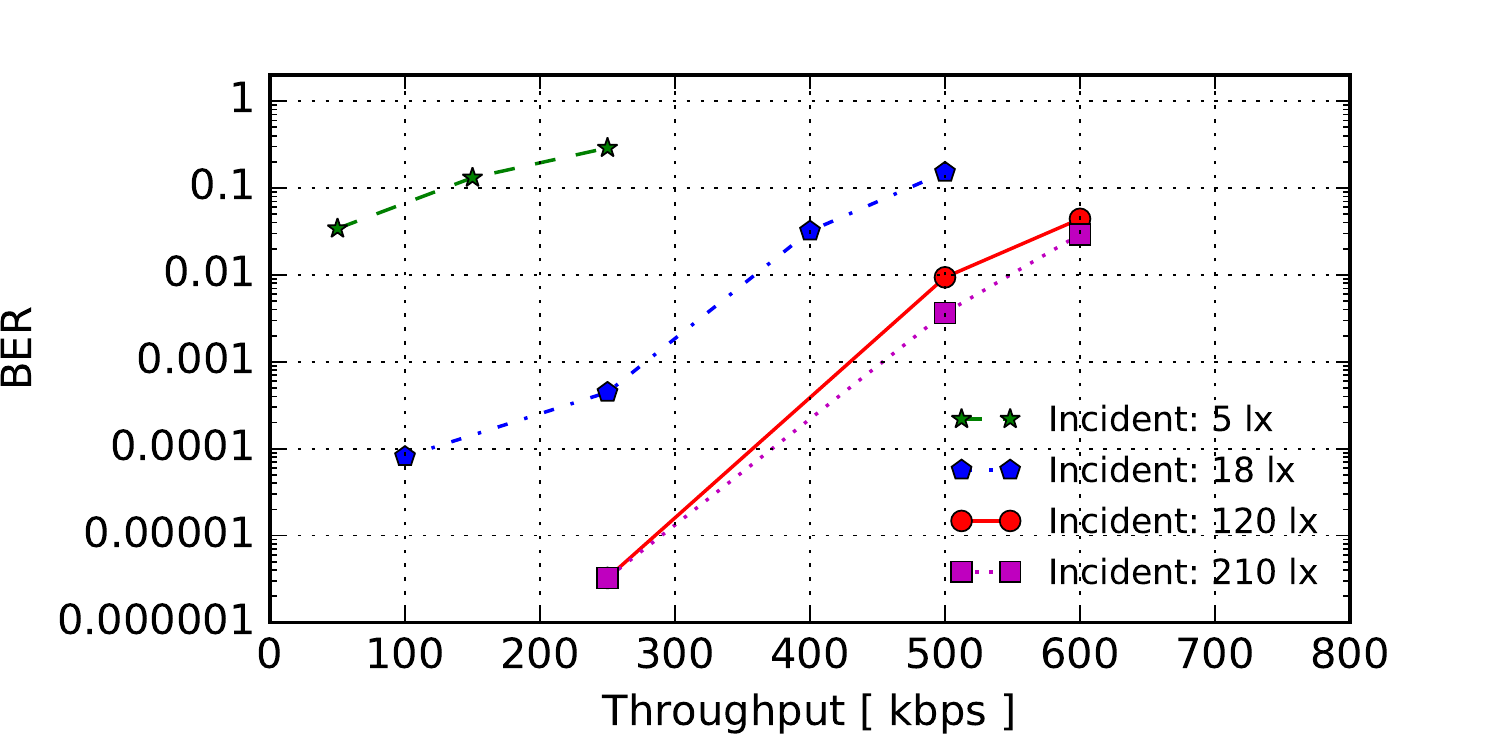}
  \caption{High gain configuration of \highspeed receiver. \capt{Due to the fixed gain bandwidth product, the receiver achieves lower throughput as compared to low-gain configuration, topping at 600 $kbps$.}}
  \label{fig:highgainhighspeed}
\end{figure}

\fakepar{Energy consumption}  Since the solar cell does not require external energy to operate,  the energy consumption of the \ulp receiver is only dictated by the power consumption of the thresholding circuit. In this circuit,
the resistor and the capacitors are passive elements and hence the comparator is the main energy consumer.
The comparator TS881 draws \SI{220}{\nano\ampere} of current resulting in a power consumption of
\SI{0.5}{\micro\watt} at \SI{2.4}{\volt}. Furthermore, the comparator enables a maximum throughput
of 500 $kbps$ which is sufficient to capture the dynamics of the solar cell.

To place these figures in perspective, we also measure the energy consumption of an actual energy harvesting device interfaced with the \ulp receiver, compared to the energy consumption of the same device when using the on-board ADC. 
We use the WISP~5.0 platform, shown in Figure~\ref{fig:wispulp}, which is powered solely by the energy harvested from RF signals.
To this end, we generate a \SI{10}{dBm} carrier signal from a software-defined radio located \SI{0.3}{\meter} away. 
Note that the distance only affects the recharge time, not the time the WISP can be active on the harvested energy. 

We modulate the VLC transmitter to send an alternating sequence of 1s  and 0s, and the WISP 
to toggle a designated GPIO pin when interrupts are received from the receiver. 
This mirrors the received bits.
We trace the GPIO output of the WISP using the logic analyzer.
We configure the LED to transmit at 2, 6, 12 and 28 $kbps$, and accordingly sample the ADC on the WISP at the same rate.
Each experiment lasts for about \SI{120}{\second} and is repeated three times. 

\begin{figure}[tb]
\centering
  \includegraphics[width=\linewidth]{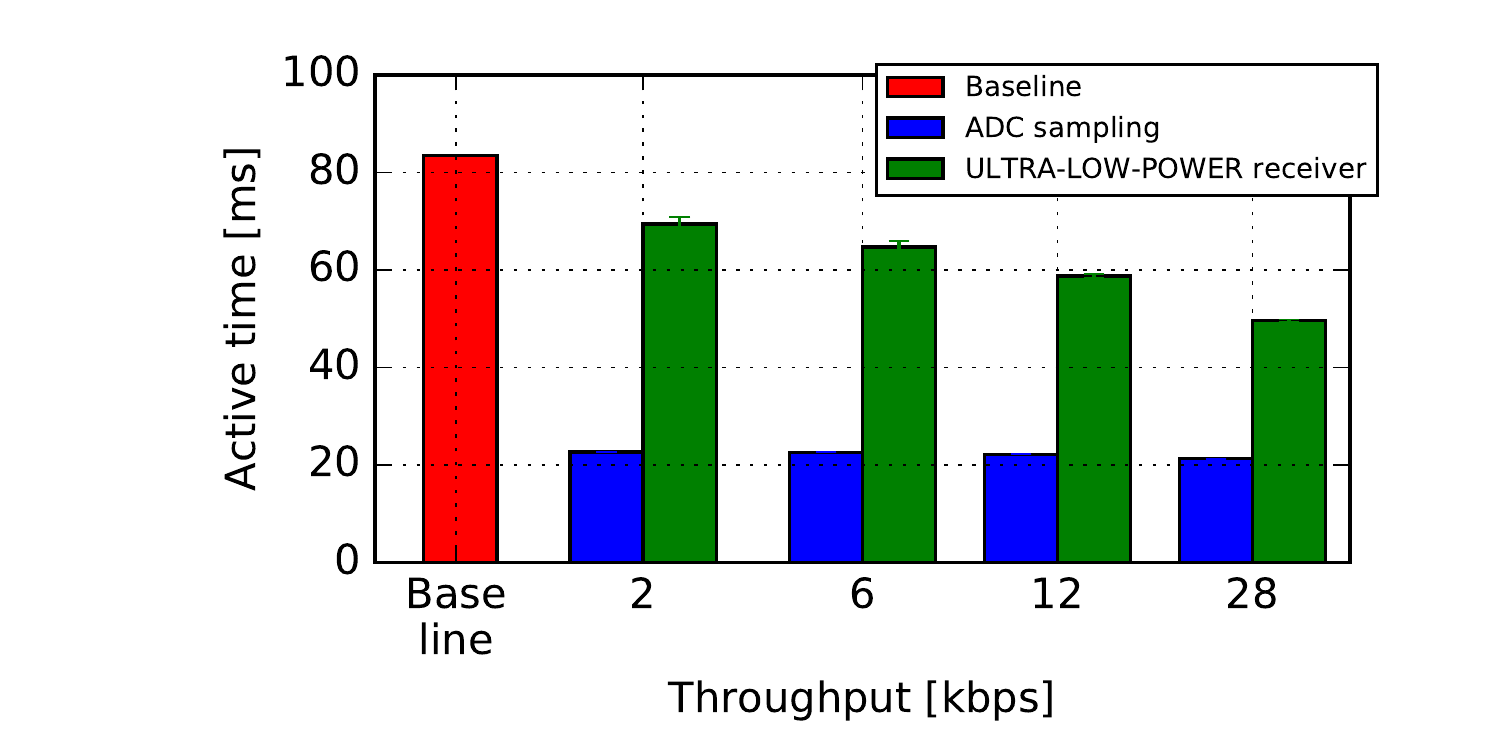}
  \caption{\ulp receiver compared to ADC sampling on WISP. \capt{Using the \ulp compared to sampling a standard photodiode through ADC operations enables improvements of a factor of two to three in active time.}}
  \label{fig:ulpreceivervsadcsamp}
\end{figure}

Figure~\ref{fig:ulpreceivervsadcsamp} demonstrates that receiving through the ADC is expensive as compared to the baseline case of not receiving at all with only the MCU active. 
The \ulp receiver performs two to three times more efficiently than sampling a photodiode. 
Lower energy consumption allows the WISP to stay active for longer, which essentially means the device can do more useful work within the same power cycle.

\begin{figure*}[!tb]
\subfigure[SNR (100 $kbps$)]{\label{fig:100kbitsnr}\includegraphics[width=0.36\linewidth]{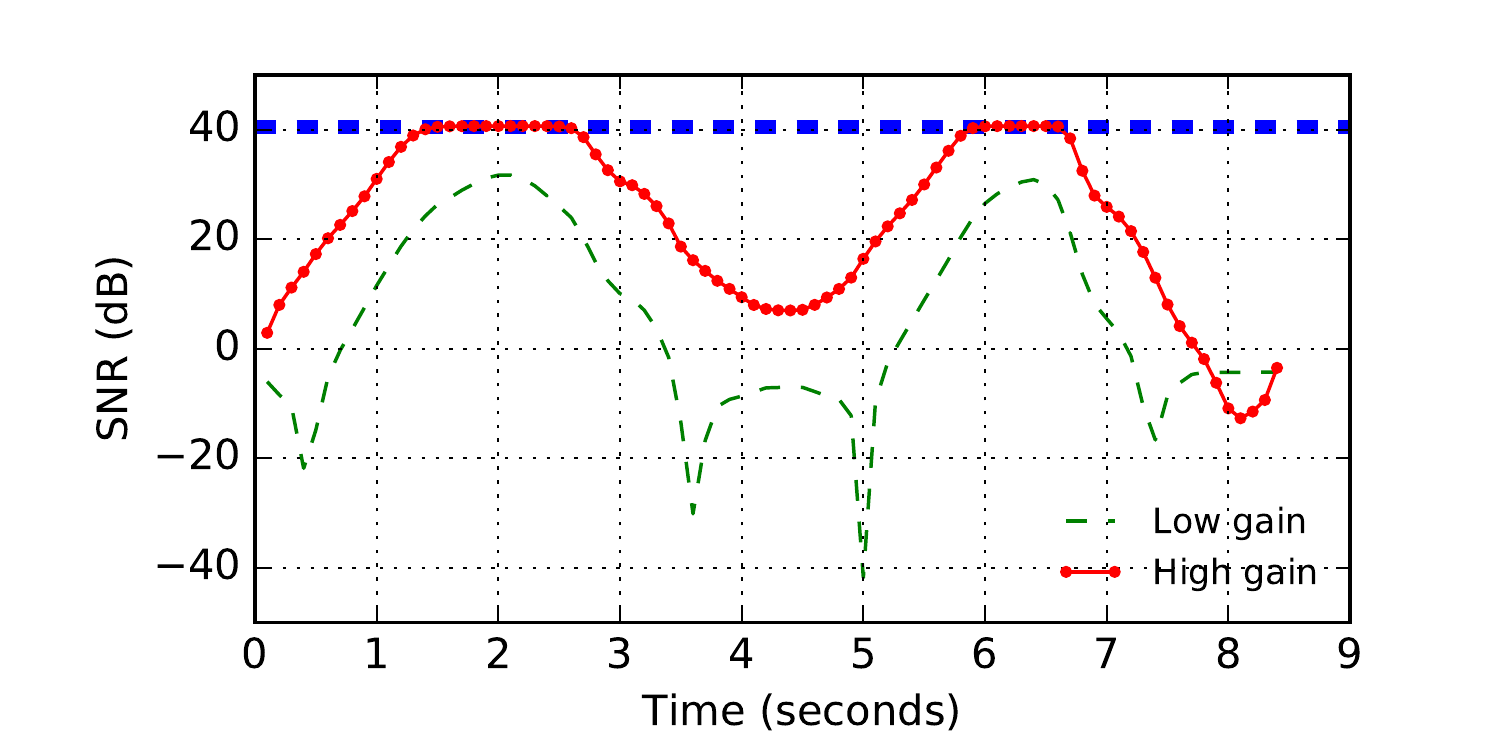}}
\hspace{-8mm}
\subfigure[BER (100 $kbps$) ]{\label{fig:100kbitber}\includegraphics[width=0.36\linewidth]{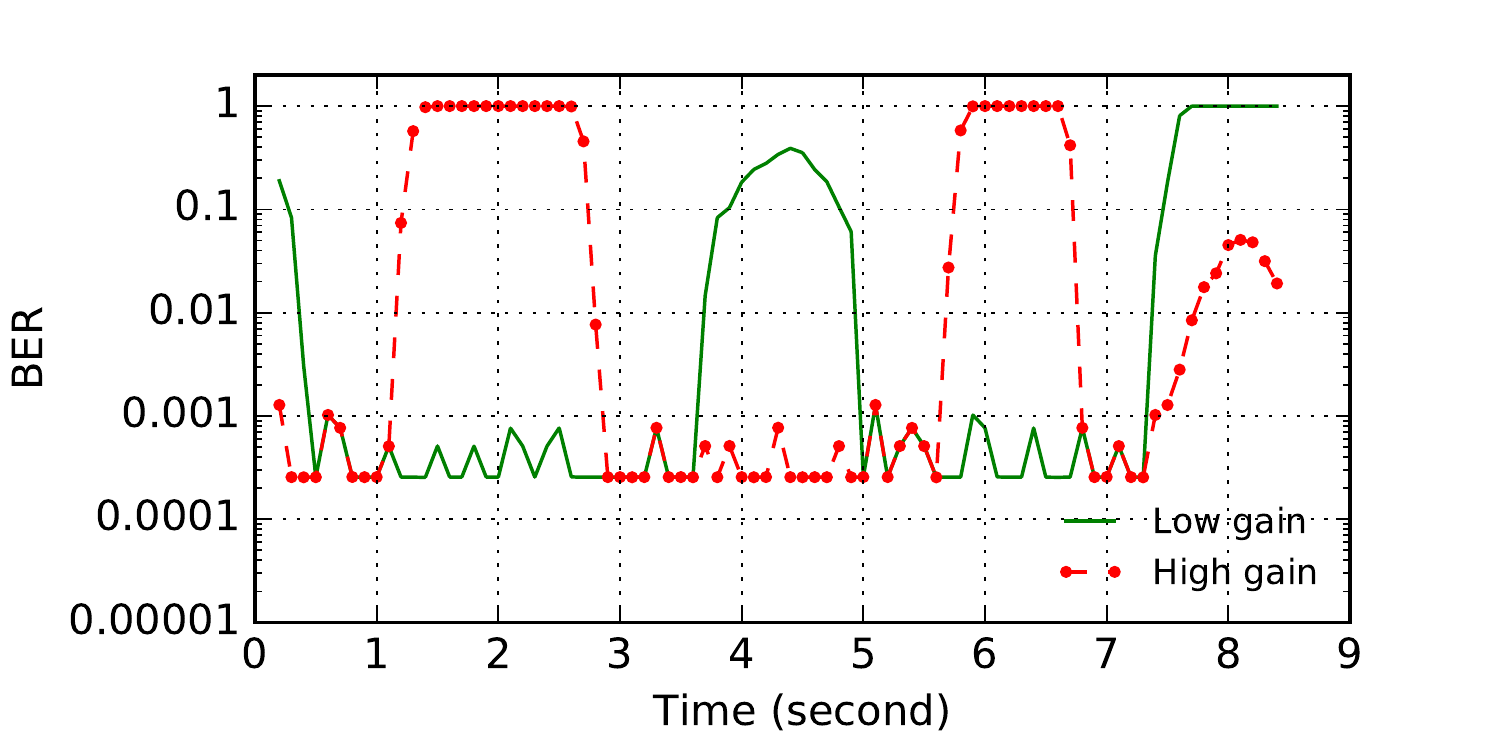}}
\hspace{-8mm}
\subfigure[Orientation BER (1000 $kbps$) ]{\label{fig:highspeedorient}\includegraphics[width=0.36\linewidth]{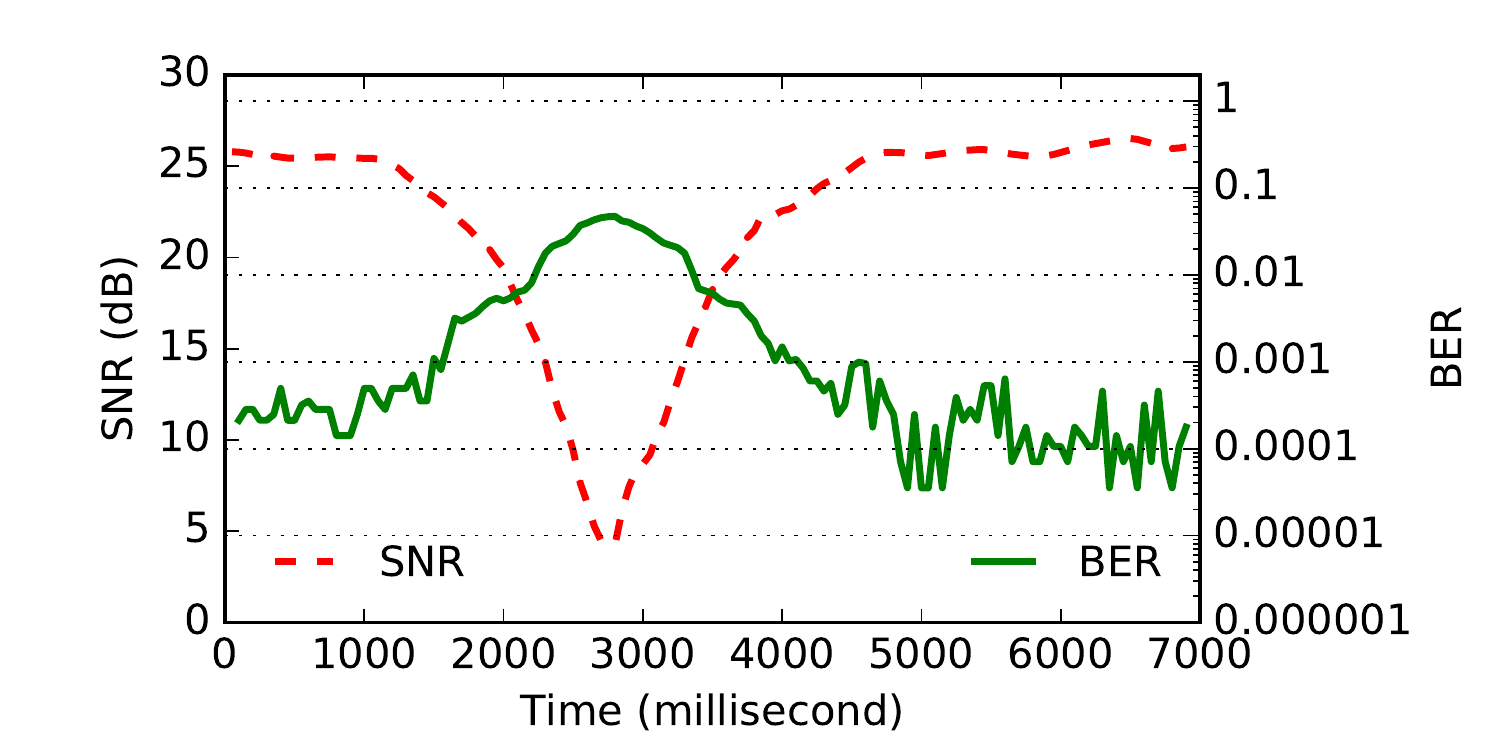}}
\vspace{-6mm}
  \caption{\highspeed receiver under mobility and orientation change. \capt{ Despite the fluctuating SNR due to mobility, the receiver does not see the BER increase as long as the SNR is positive. The dashed blue line indicates the level at which the high speed receiver saturates. }}
\label{fig:errors_dynamics_highspeed}
\end{figure*}

\subsection{\highspeed Receiver}  
\label{sec:eval-high}

\begin{figure*}[!tb]
\centering
\subfigure[BER (100 $kbps$) \label{100kbp_snr}]{\includegraphics[width=0.23\linewidth]{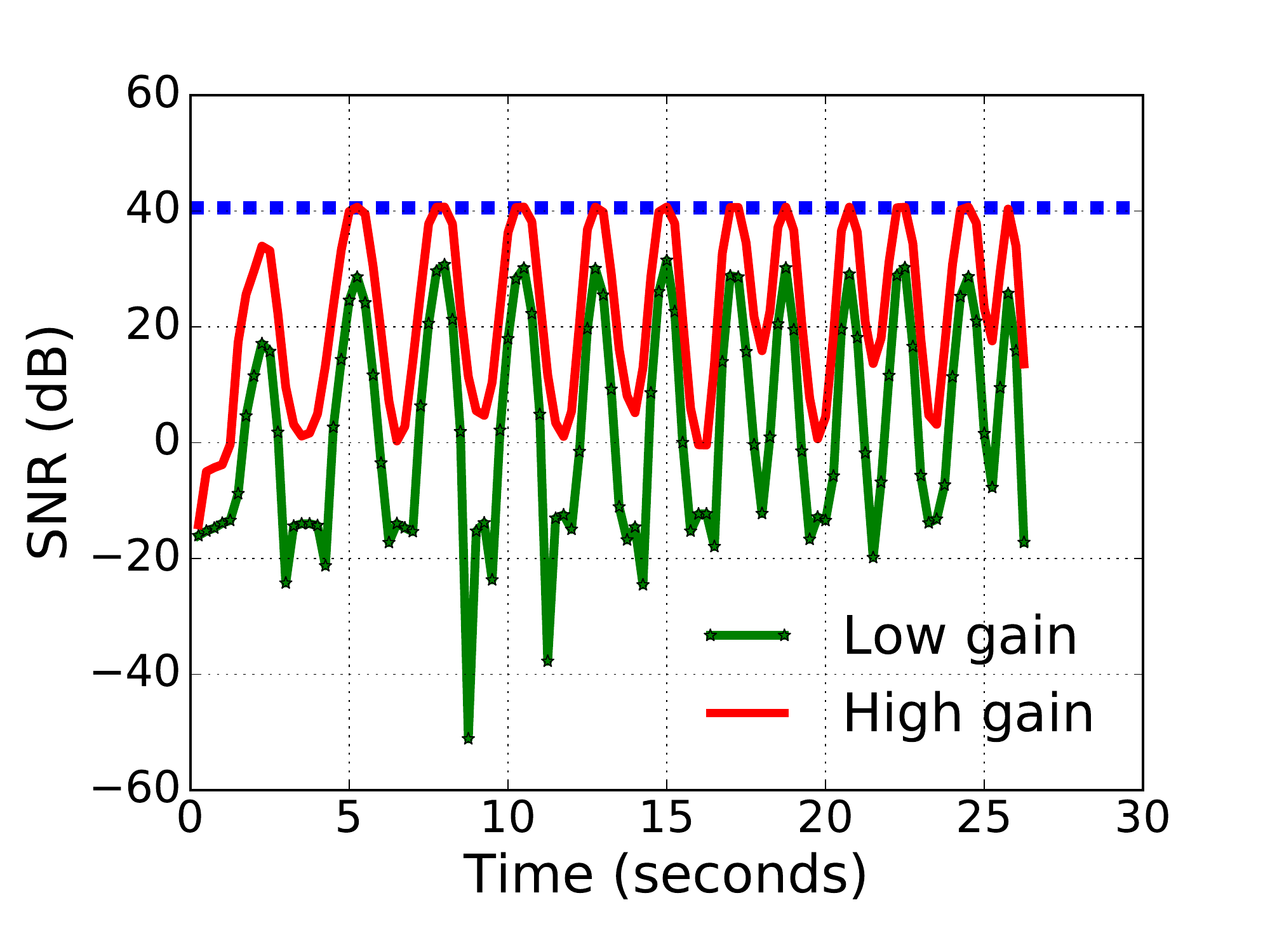}}\quad
\subfigure[Low gain~(100 $kbps$)\label{100kbps_lowgain}]{ \includegraphics[width=0.23\linewidth]{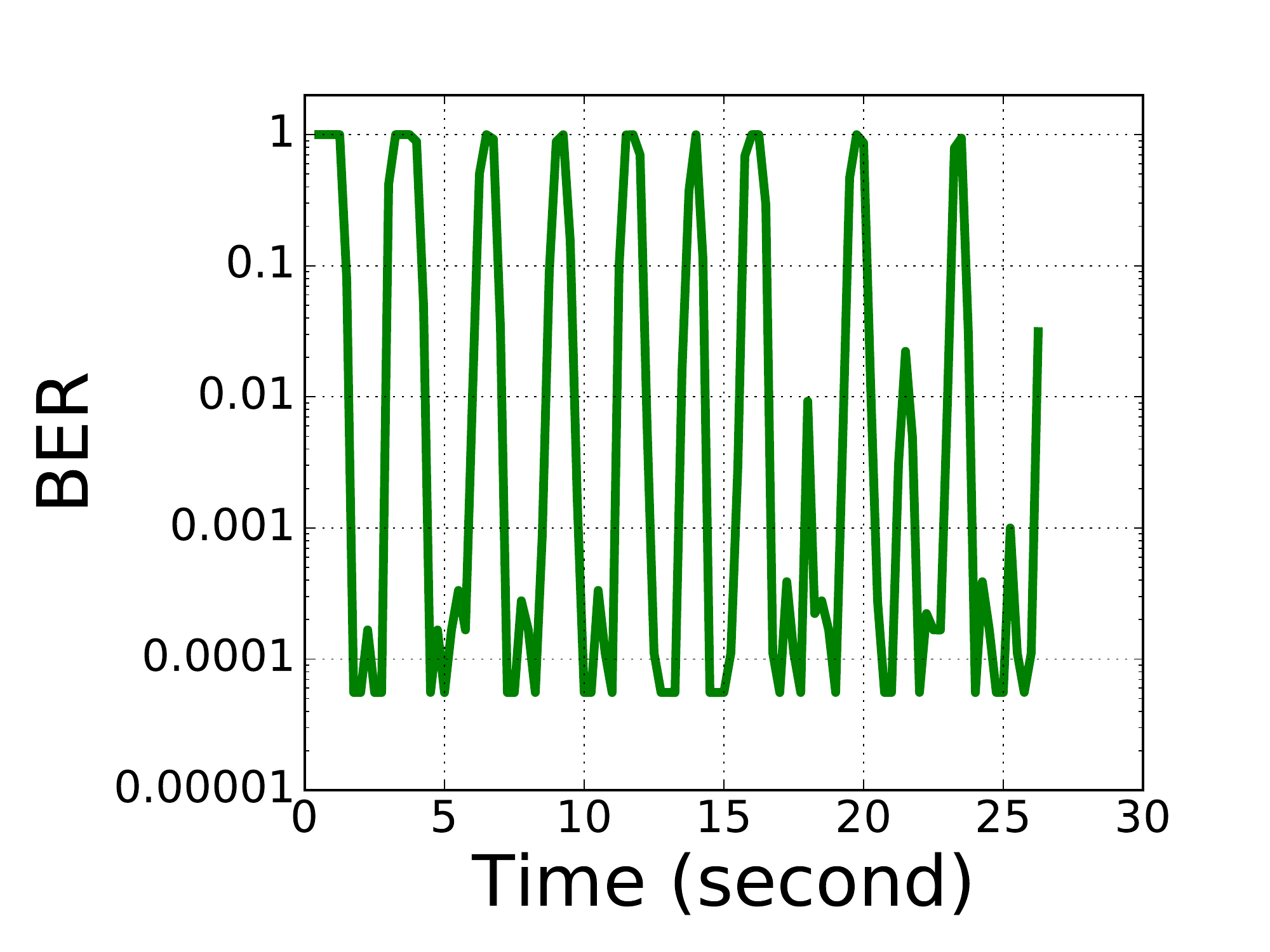}}\quad
\subfigure[High gain~(100 $kbps$)\label{100kbps_highgain}]{ \includegraphics[width=0.23\linewidth]{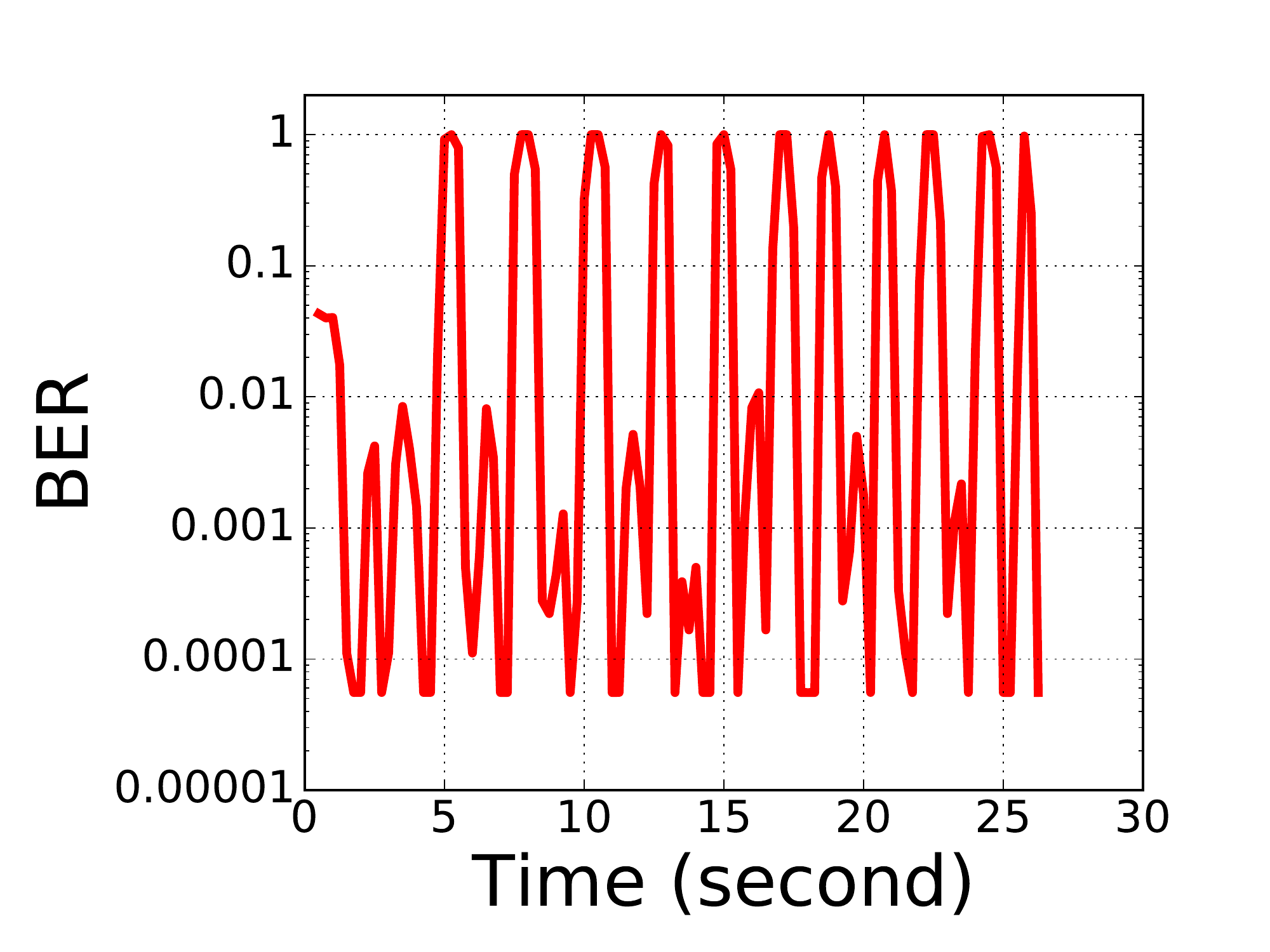}}\quad
\subfigure[Switching logic~(100 $kbps$)\label{100kbps_switch}]{ \includegraphics[width=0.23\linewidth]{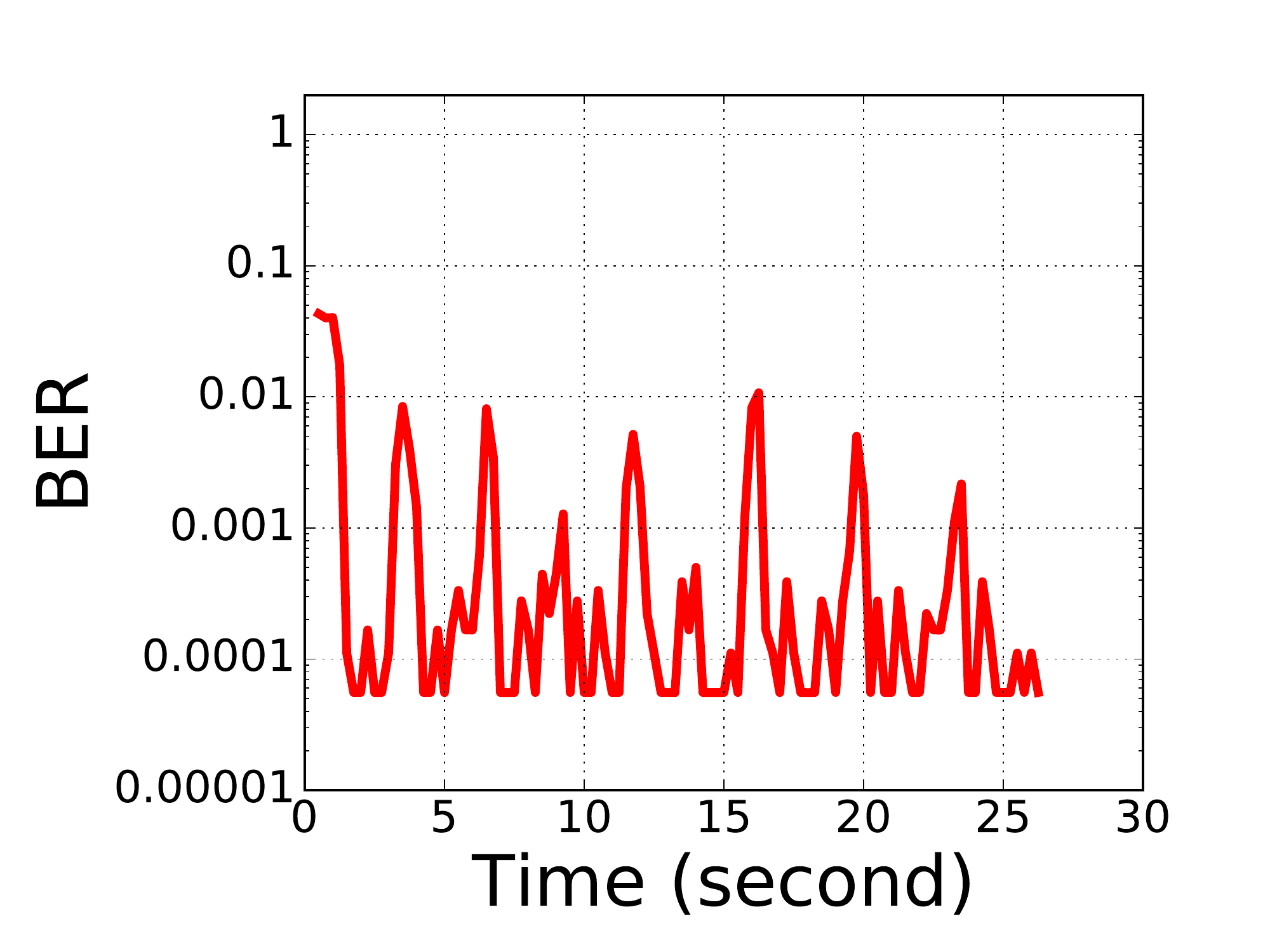}}\quad
\vspace{-6mm}
\caption{BER observed as \platform unit adapts to changing light conditions. \capt{When the operating light conditions change, the \platform unit switches
  between the two gain configurations of the \highspeed receiver. Even though the performance of the individual receivers vary wildly, the BER remains low.}}
  \vspace{-6mm}

\label{switching:throughput}
\end{figure*}

We evaluate to what extent the \highspeed receiver can fulfill its main design goal of high throughput. 

\fakepar{Low gain Throughput} 
In this experiment, we measure the throughput achieved by the low gain
configuration of the \highspeed receiver in different ambient light conditions and incident light levels, as a function of the transmitter's data rate.
Figure~\ref{receiver:highspeed} reports the results.
Unlike the results achieved with the \ulp receiver shown in 
Figure~\ref{ulp:receivereval}, here the incident light levels significantly affect the achievable throughput: at high incident light levels, the SNR increases which leads to much better performance.

Figure~\ref{receiver:darkness} shows that, for example, the \highspeed receiver achieves a throughput of 1700 $kbps$ in the absence 
of ambient light and with incident light levels of \SI{925}{\lux}. This throughput is comparable with that of WiFi~(IEEE 802.11b) and Bluetooth chips.
The BER rises sharply after this point, as we reach the operational limit of the thresholding circuit.
At low incident light levels, the maximum achievable throughput is slightly above 1000 $kbps$.  
Similar trends are found in different ambient light conditions, as shown in Figure~\ref{receiver:officelight} and~\ref{receiver:naturallight}.
The trends are independent of the ambient light conditions since
the ambient light only adds a DC offset to the signal. This offset is, however, averaged out by the thresholding circuit we employ in the \highspeed receiver as described in Section~\ref{sec:hs}. 

\fakepar{High gain throughput} Next, we measure the throughput achieved by the high gain configuration of the \highspeed receiver. As the TIA has a fixed gain bandwidth product, we expect that a higher gain lowers the achievable throughput. Due to the high gain, the receiver saturates even at moderate light conditions like in the natural lighting or indoor lighting settings. Hence, we perform the experiment in darkness. Figure~\ref{fig:highgainhighspeed} demonstrates the result of the experiment. The figure shows that the receiver achieves a maximum throughput of 500 $kbps$ at a BER of $10^{-3}$.  Furthermore, impressively the receiver achieves a high throughput of 500 $kbps$ even at a very low light intensity level of \SI{18}{\lux}.

\fakepar{Sensitivity} Table~\ref{sensitvitytable} shows the sensitivity 
of the two gain configurations of the \highspeed receiver. 
As expected, the high gain configuration can operate at significantly
low light levels. At extremely low light levels corresponding to 
\SI{4}{\lux}, the receiver 
can still support a throughput of 100 $kbps$. As the light levels increase slightly, the
receiver can support a throughput as high as 500 $kbps$ at light levels of \SI{25}{\lux}. On the
other hand, the low gain configuration of the receiver starts to operate at light
levels of  \SI{25}{\lux} and achieves a throughput of 100 $kbps$.

\fakepar{Mobility}  We evaluate the \highspeed receiver under mobility performing an experiment similar to the one with the \ulp receiver. In contrast to the earlier experiment, 
we operate the VLC transmitter at a data rate of 100 $kbps$, 
the maximum throughput supported by the high gain receiver at very low light conditions~(see Table~\ref{sensitvitytable}).
We induce changes in the light intensity from a minimum of
\SI{6}{\lux} to \SI{800}{\lux} at an interval of \SI{8}{\second}.
Figure~\ref{fig:100kbitsnr} shows the resulting SNR as observed by the two
\highspeed receivers.

Figure~\ref{fig:100kbitber} shows that the BER on individual receivers varies between 0 indicating no
error to 1 indicating all bits were lost as the light intensity levels fluctuate. The low gain configuration fails to detect transmissions at lower light levels, while the high gain configuration gets saturated
easily at moderate light levels and hence fails to detect transmissions. We note, when the high gain receiver 
fails to operate due to saturation, the low gain receiver continues to operate at a low BER. 
On the other hand, when the low gain configuration fails to operate due to a negative SNR, 
the high gain receiver is able to detect transmissions.
We precisely use this fact in the \platform unit
to tackle the varying light intensity levels caused by mobility.

\fakepar{Orientation change}  In the next experiment, we evaluate the \highspeed receiver  when the orientation changes, similar
to the experiment conducted earlier for the \ulp receiver.
We configure the VLC transmitter to send at a data rate of 1 $Mbps$ which falls in the operating range of the
low gain receiver. Next, we change the orientation of the receiver varying its angle from the 
transmitter from 0 (straight up) to  90 degrees and back to 0 degrees. Figure~\ref{fig:highspeedorient}
demonstrates that the SNR can change by almost 20 $dB$ due to the
change in orientation. But even under these conditions the receiver 
maintains a low BER that is mostly below $10^{-3}$.

\fakepar{Energy consumption} Throughout the experiments we carry out, the \highspeed receiver shows an energy consumption comparable with that of WiFi and Bluetooth chips for the same throughput, as shown in Figure~\ref{fig:highspeedenergybit}. Even though optimizing this metric was not a design goal for the \highspeed receiver, the much higher throughput we obtain compared to the state-of-the-art in embedded VLC is not detrimental to the energy consumption.

\subsection{\platform Unit}  
\label{sec:evalss}

The switching logic on the \platform unit selects the appropriate receiver for a given light condition and performance goal. Furthermore, it is possible to interface
the board with the wearable device. 
In this section, we evaluate the performance of the
switching logic and its integration with the wearable host device.

\fakepar{Optimising throughput} First, we evaluate the performance of the switching logic when maximizing the throughput. We program the VLC transmitter to send at a data rate of 100~$kbps$, which is the maximum throughput 
of the high gain configuration of the \highspeed receiver at low light levels.
We alter the light intensity levels by changing the orientation
of the bulb rapidly such that the light levels switch between \SI{3}{\lux}
and \SI{1500}{\lux} 12 times within \SI{27}{\second} as 
shown in Figure~\ref{100kbp_snr}. We note, the change of the light levels within the short time 
span we induce are extreme compared to changes that can be expected 
under mobility~\cite{zhang2015dancing}, and hence represent a worst case
scenario.  

Figure~\ref{100kbps_lowgain} and~\ref{100kbps_highgain} show the BER calculated
with a moving window of \SI{250}{\milli\second}. The BER of the individual receivers
varies greatly. The BER is 1, that is no bits are received, during low-light conditions or under saturation, while it is low in the receivers' operating regions. 
Figure~\ref{100kbps_switch} shows that the switching logic appropriately
selects the best performing receiver, and hence is able to keep the BER 
low for the   whole duration of the experiment.

\begin{figure}[!tb]
\centering
  \includegraphics[width=0.4\textwidth]{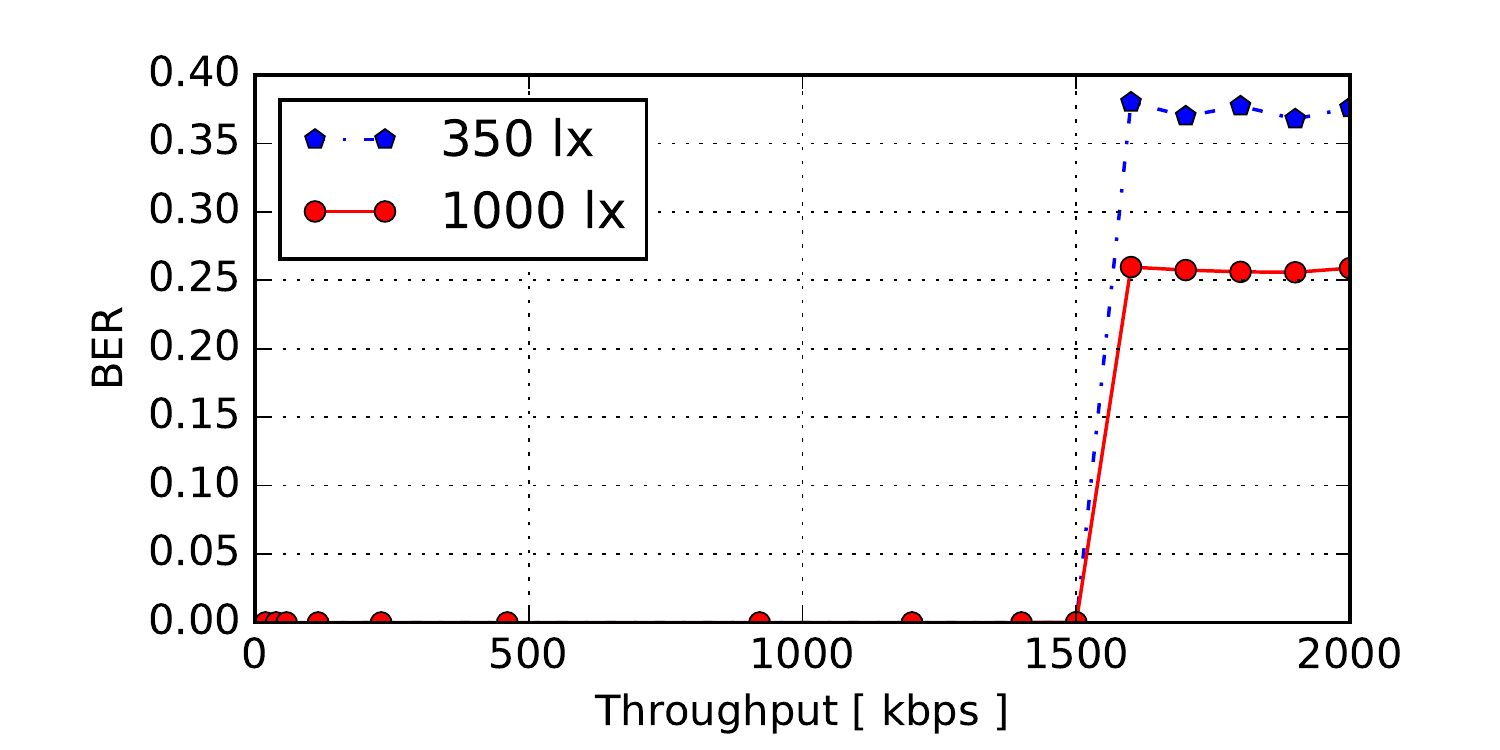}
  \caption{Throughput and BER of the \platform unit using the \highspeed receiver and the UART-to-USB converter. \capt{We support a maximum throughput of 1500 $KBit/s$ with 0 BER. } }
  \label{uarteval}
\end{figure}

\fakepar{Optimising energy per bit}  Periodically probing the channel to determine
the light conditions can be energy expensive due to the use of an ADC.
To resolve this, in the switching logic we sample only at a 
fixed interval $\Delta$. In this experiment we determine how the choice
of $\Delta$ effects the response time, that is, the time the receiver needs to respond to changed light conditions.

\begin{figure}[!tb]
\centering
  \includegraphics[width=0.4\textwidth]{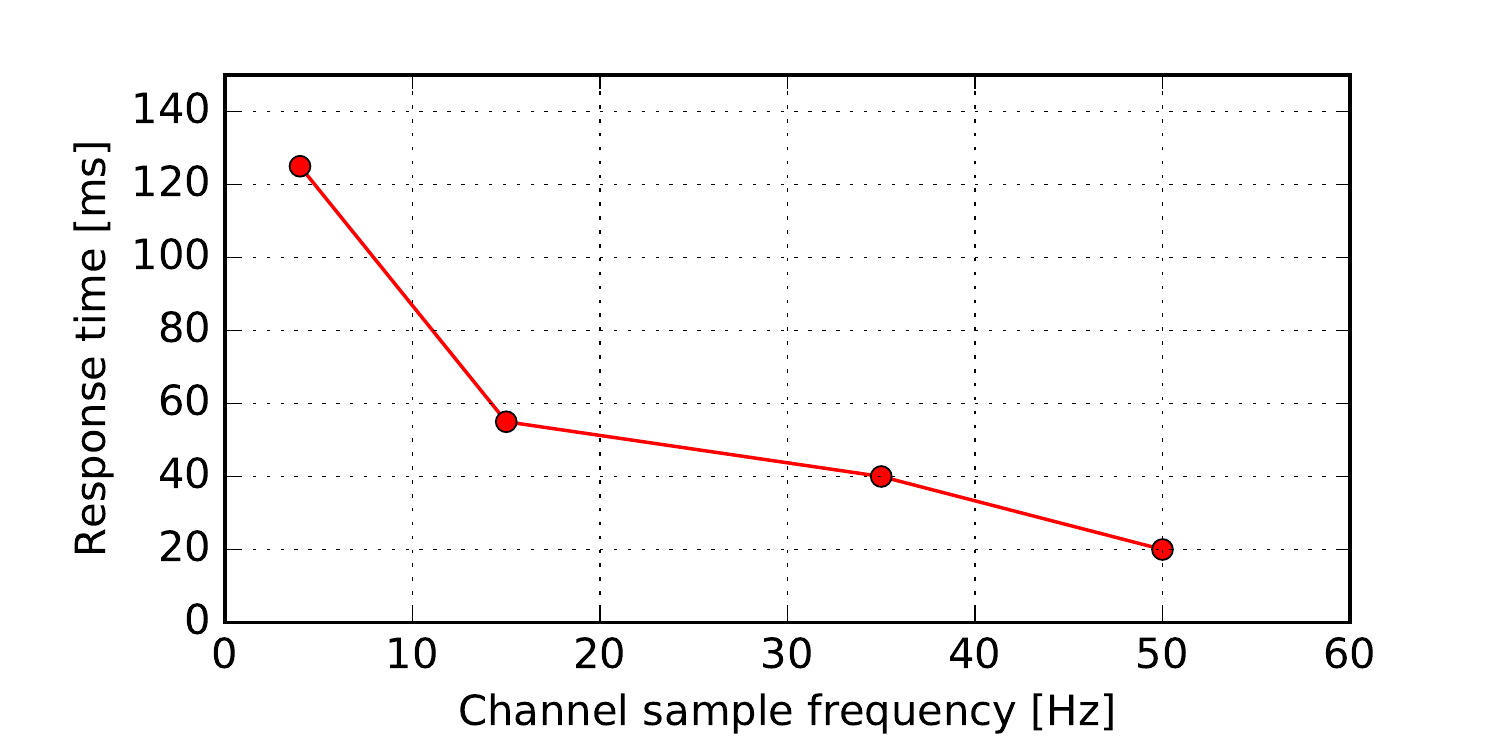}
  \caption{Response time to change in channel conditions \capt{A higher sampling frequency results in lower response time at the expense of increased energy consumption. We use \SI{50}{\hertz} sampling rate to optimise throughput.} }
  \label{responsetime}
\end{figure}

Similar to the earlier experiment involving mobility, we change the light intensity
on the receiver. We program the switching logic to sample at frequencies of
4, 15,35 and \SI{50}{\hertz}, which are low enough to ensure low energy consumption.
We record the time it takes the receiver
to switch to the high gain configuration when the light intensity decreases. Figure~\ref{responsetime}
demonstrates that the response time, as expected, decreases with an increasing sampling frequency, and reaches
\SI{20}{\milli\second} for $\Delta$ of \SI{50}{\hertz}. At low sampling frequencies it takes a significant amount of time for 
the receiver to respond and switch to the best performing receiver.

\fakepar{Throughput} We first evaluate the throughput when the wearable device is interfaced using the USB controller chip. 
In principle, we expect the performance to be similar to that of \highspeed receiver discussed earlier.
However, Figure~\ref{uarteval} demonstrates that the 
\platform unit tops at 1500 $kbps$ with zero BER.

There are two reasons for this performance.
First, UART communication adds an overhead in terms of parity and stop bits.
Second, as we trace the output of the UART, we process bytes and not bits as in the previous experiments.
Hence, even a single incorrectly received bit makes a whole byte erroneous.
Together these two factors lead to a lower throughput compared to a stand-alone 
\highspeed receiver.

On the low-power MCU on the \platform unit, we are able to demodulate at a maximum throughput of 1200 $kbps$, when the MCU is operating at
24 $MHz$ of operating clock frequency.  We expect the achieved throughput to be similar for other wearable platforms  employing similar low-power
MCUs when interfaced to the \highspeed receiver.

%% file: end2.tex
\section{Conclusion}
\label{sec:end}

VLC in mobile settings is  extremely challenging 
since mobility induces drastic changes in the SNR. 
We present three different 
VLC receiver designs. While outperforming
state-of-the-art VLC receivers none of our
individual receivers alone can cope with the rapid SNR changes
in mobile settings. Hence, we design a switching logic that
we implement on an integration unit. Our experiments show that 
our logic is able to rapidly adapt to the fluctuating light conditions and
select the best performing receiver.    
